\documentclass[twocolumn,twocolappendix]{EBs_study}

\begin{document}

\title{Comparative Analysis of Ellerman and Quiet Sun Ellerman Bombs in the Solar Atmosphere}

\author[orcid=0009-0002-0549-1446,gname= Ravi,sname='Chaurasiya']{Ravi Chaurasiya}
\altaffiliation{Indian Institute of Technology, Gandhinagar, Gujarat-382355, India}
\affiliation{Udaipur Solar Observatory, Physical Research Laboratory, Udaipur-313001, India}
\email[show]{ravi@prl.res.in}  

\author[orcid=0000-0001-5802-7677,gname=Ankala, sname='Raja Bayanna']{Ankala Raja Bayanna} 
\affiliation{Udaipur Solar Observatory, Physical Research Laboratory, Udaipur-313001, India}
\email{bayanna@prl.res.in}

\author[orcid=0000-0003-0585-7030,gname=Jayant,sname=Joshi]{Jayant Joshi}
\affiliation{Indian Institute of Astrophysics, II Block, Koramangala, Bengaluru 560 034, India}
\email{jayant.joshi@iiap.res.in}

\begin{abstract}
Ellerman Bombs (EBs) and Quiet-Sun Ellerman Bombs (QSEBs) are small-scale signatures of magnetic reconnection in the lower solar atmosphere, observed in active regions and quiet-Sun areas, respectively. We investigate and compare some of their properties using coordinated multiwavelength observations from the Swedish 1-m Solar Telescope, the Interface Region Imaging Spectrograph, and the Solar Dynamics Observatory. We employ k-means clustering to identify EBs and QSEBs and perform a detailed analysis of a subset of these events. Our results show that EBs are frequently associated with opposite magnetic polarities, whereas QSEBs generally lack clear polarity signatures, likely due to limited spatial resolution. Spectral inversions using the STiC code reveal temperature enhancements of up to $\sim$1700 K in the lower chromosphere for EBs. In contrast, no clear temperature enhancement is detected for QSEBs, which may be attributed to the limited spatial resolution or insufficient wavelength sampling of the Ca\,\textsc{ii} 8542~\AA. We further find that some EBs exhibit signatures extending to transition-region temperatures. An analysis of EBs temporal evolution reveals episodic heating, with a range of periodicities, most commonly around 6–7 minutes. In addition, we identify spatial associations between the footpoints of some spicules and EBs/QSEBs, suggesting that reconnection in these events may contribute to spicule formation. These results demonstrate similarities and differences between EBs and QSEBs and support the interpretation that small-scale magnetic reconnection contributes to heating and dynamics in EBs, while the underlying mechanism of QSEBs requires further investigation.
\end{abstract}


\section{Introduction} 

The solar atmosphere is a highly dynamic, magnetized plasma environment composed of multiple layers namely photosphere, chromosphere, transition region, and corona, each characterized by distinct temperature and density profiles. This complex system hosts a wide range of phenomena resulting from the interplay of magnetic fields, plasma dynamics, and radiative processes, including numerous jet-like structures such as spicules \citep{2004Natur.430..536D,2014ApJ...792L..15P,2019ARA&A..57..189C,Bose.et.al.21,2022NatPh..18..595D,2024ApJ...970..179C}, and pervasive wave activity across atmospheric layers \citep{2009SSRv..149..355Z,2013SSRv..175....1M,2015SSRv..190..103J,2023LRSP...20....1J}.

\begin{figure*}
	\centering
	\includegraphics[width=180mm]{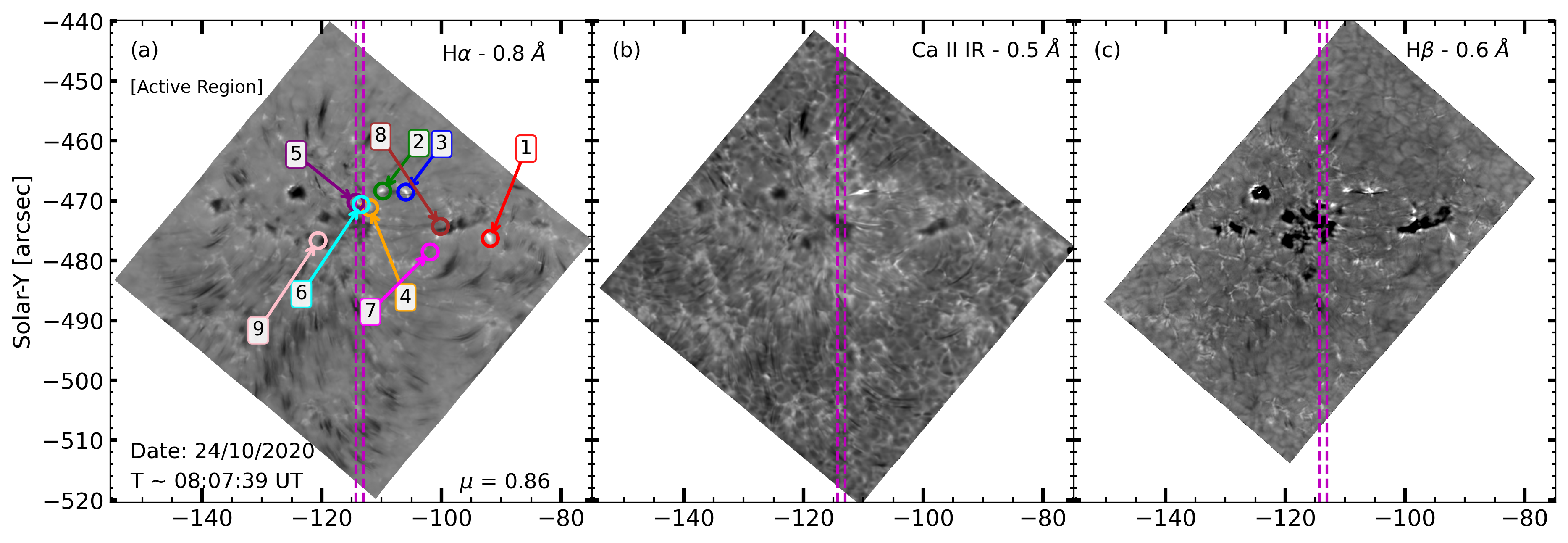}
	\includegraphics[width=180mm]{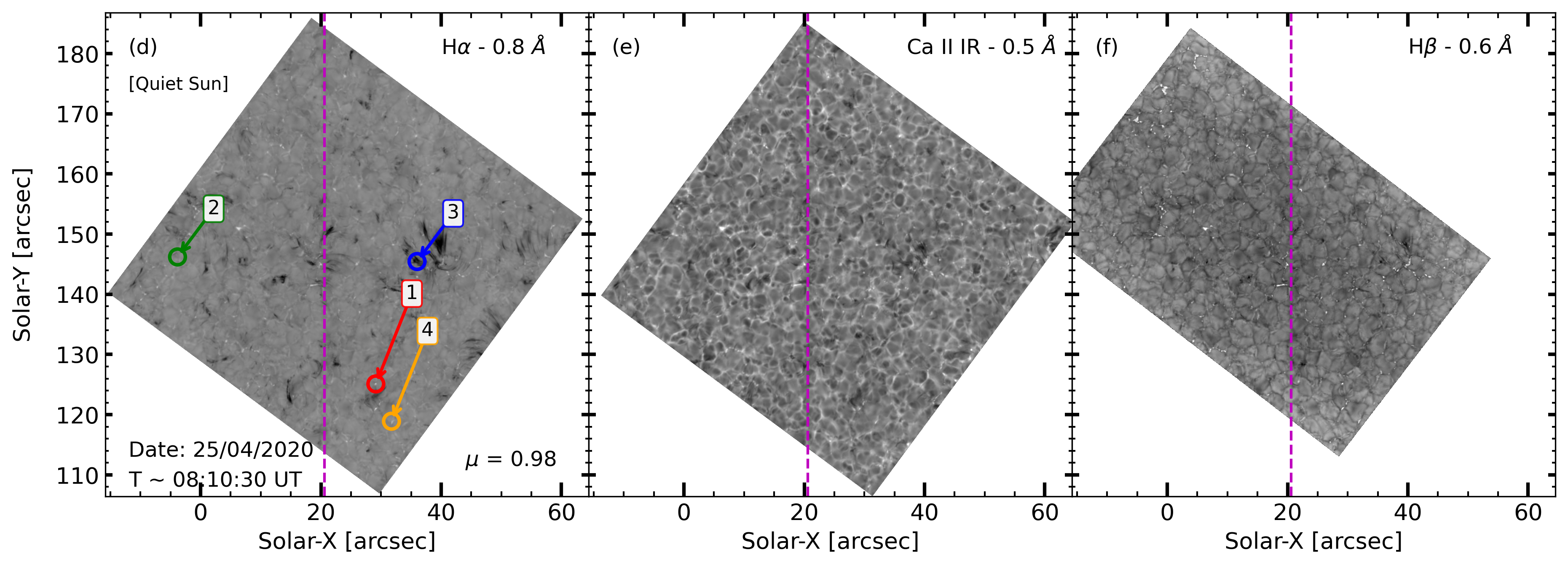}
	\caption{The upper panel shows the chromosphere of the active region (Dataset 1) as observed in the blue wings of H$\alpha$ (H$\alpha$~$-0.8$\,\AA), Ca\,\textsc{ii} IR (Ca\,\textsc{ii} IR~$-0.5$\,\AA), and H$\beta$ (H$\beta$~$-0.6$\,\AA). The lower panel presents the quiet Sun (Dataset 2) chromosphere observed in the same spectral line wings. The magenta vertical line indicates the IRIS slit coverage. The colored labels in both the upper and lower panels mark the locations of the EBs and QSEBs studied in this work in detail.}
	
	\label{Fig1}
\end{figure*}

In addition to these wave and jet-like dynamics, the chromosphere also hosts transient, small-scale energetic events known as Ellerman Bombs (EBs). EBs are characterized by small scale short lived localized brightenings in the wings of the Balmer lines and Ca\,\textsc{ii} 8542~\AA{} lines \citep{1917ApJ....46..298E,2002ApJ...575..506G,2011ApJ...736...71W,2013JPhCS.440a2007R,2013ApJ...774...32V,2015ApJ...812...11V}. Typically found in emerging active regions with strong magnetic flux, EBs appear near polarity inversion lines as sub-arcsecond bright structures in the line wings of the chromospheric spectral lines \citep{2002ApJ...575..506G,Pariat.et.al.07,2013A&A...557A.102B,2017A&A...598A..33L,2017GSL.....4...30C}. The characteristic signature of an EB is the strong enhancement in the line wings without corresponding brightening in the core, giving rise to the characteristic ``moustache'' profile in H$\alpha$. When observed at an inclined angle, these events resemble small (1–2~Mm), vertical flame-like structures that show flickering on timescales of a few seconds \citep{2011ApJ...736...71W,2013JPhCS.440a2007R,2015ApJ...798...19N}.

High-resolution imaging spectroscopy has revealed that the enhanced wing emission arises due to localized heating in the lower atmosphere, while the overlying chromospheric fibril canopy conceals the reconnection site \citep{2011ApJ...736...71W,2013JPhCS.440a2007R,2013ApJ...779..125N,2015ApJ...812...11V,2019ARA&A..57..189C}. This interpretation, which places EBs beneath the chromospheric canopy, is further supported by recent observations from the Atacama Large Millimeter/submillimeter Array (ALMA) \citep{2020A&A...643A..41D} and state-of-the-art radiative magnetohydrodynamic simulations \citep{2017ApJ...839...22H,2019A&A...626A..33H,2017A&A...601A.122D}. In particular, \citet{2019A&A...626A..33H} investigated the connection between EBs and their ultraviolet (UV) counterparts (IRIS UV burst-type emissions (e.g., \text{Si\,{\sc iv}} signatures) that are associated with a subset of EBs \citep{2018SSRv..214..120Y}). Based on radiative MHD simulations, they suggested that both phenomena may originate from a vertically extended current sheet. In these simulations, the current sheet extends up to $\sim$ 3 Mm above the photosphere; however, this extent depends on the adopted model assumptions and initial conditions and may therefore differ in the real solar atmosphere.

Expanding the scope of EB-like phenomena, \citet{2016A&A...592A.100R} found that EB-like phenomena are not just limited to the active regions and identified similar brightenings outside active regions, termed Quiet Sun Ellerman-like Brightenings (QSEBs), using H$\alpha$ observations from the Swedish 1-m Solar Telescope (SST; \citealt{2003SPIE.4853..341S}). Follow-up studies using H$\beta$ observations, which offer improved spatial resolution and contrast due to their shorter wavelength, have revealed the widespread presence of QSEBs in the quiet Sun. \citet{2020A&A...641L...5J} estimated that nearly 500{,}000 QSEBs may exist across the solar disk at any time, a number later revised to approximately 750{,}000 by \citet{2024A&A...683A.190R} using higher-resolution H$\epsilon$ observations. \cite{2024A&A...689A.156B} investigated 453 long-lived QSEBs ($>$ 1 min) and found 67 cases exhibiting co-spatial and co-temporal UV brightenings in IRIS SJI 1400 \AA, suggesting that a subset of QSEBs may be heated to transition-region temperatures. Additionally, the origin of spicules has been linked to QSEBs (\citealt{2025A&A...697A.180S}). These results highlight the ubiquity of QSEBs and their potential significance in localized energy transport in the lower solar atmosphere.

In this study, we investigate and compare the properties of EBs and QSEBs using coordinated multiwavelength observations from the SST \citep{2003SPIE.4853..341S}, Interface Region Imaging Spectrograph (IRIS; \cite{2014SoPh..289.2733D}), and (Solar Dynamics Observatory (SDO); \citep{2012SoPh..275....3P})/Atmospheric Imaging Assembly (AIA, \cite{2012SoPh..275...17L}) and SDO/Helioseismic Magnetic Imager (HMI,\space \cite{2012SoPh..275..229S}). Additionally, we examine the potential connection between EBs/QSEBs and the origin of some spicular structures in the solar atmosphere.

The remainder of the paper is organized as follows: Section~\ref{sec2} outlines the data reduction and alignment procedures. Section~\ref{sec3} presents our observational results, analysis and discussion. Finally, summary and conclusions are drawn in Section~\ref{sec5}.

\begin{table*}
	\centering
	\caption{Summary of SST, IRIS, and SDO observations used in this study.}
	\label{tab1}
	
	\vspace{0.5em}
	\textbf{(a) SST Observations. \\ Format: Average Cadence (s) / \#Positions / Range (\AA) around the line center.} \\
	\begin{tabular}{lccccc}
		\hline
		\textbf{Spectral Line} & H$\alpha$ & Ca\,\textsc{ii} 8542 & Fe\,\textsc{i} 6302/6173 & H$\beta$ & Ca\,\textsc{ii} H \& K \\
		\hline
		Dataset 1 (AR) & 34.25/17/$\pm$2.0 & 34.25/15/$\pm$1.3 & 34.25/15/$\pm$0.34 & 17.59/17/$\pm$1.6 & 17.31/31/$\pm$2.0 \\
		Dataset 2 (QS) & 10.50/7/$\pm$1.3 & 10.50/7/$\pm$1.0 & 10.50/7/$\pm$0.30 & 16.58/17/$\pm$1.6 & 16.62/31/$\pm$2.0 \\
		\hline
	\end{tabular}
	
	\vspace{1em}
	\textbf{(b) IRIS Observations} \\
	\begin{tabular}{llll}
		\hline
		\textbf{Parameter} & \textbf{Dataset 1} & \textbf{Dataset 2} \\
		\hline
		\multicolumn{3}{c}{\textit{Slit-Jaw Images (SJIs)}} \\
		Channels & \multicolumn{2}{c}{1400\,\AA\ (Si\,\textsc{iv}), 2796\,\AA\ (Mg\,\textsc{ii}\,k), 2832\,\AA\ (Mg\,\textsc{ii}\,h wing)} \\
		Plate scale & 0.3327$''$ & 0.1664$''$ \\
		Cadence & 1400\,\AA: 18.22\,s & 1400\,\AA: 28.68\,s  \\
		& 2796\,\AA: 18.22\,s & 2796\,\AA: 28.68\,s  \\
		& 2832\,\AA: 109.35\,s & 2832\,\AA: 28.68\,s  \\
		\hline
		\multicolumn{3}{c}{\textit{Spectroscopy}} \\
        Observed lines & \multicolumn{2}{c}{C\,\textsc{ii} 1336\,\AA, Si\,\textsc{iv} 1394/1403\,\AA, Mg\,\textsc{ii}\,k 2796\,\AA} \\
		Raster mode & Large dense 4-step raster  & Large sit-and-stare mode  \\
		Cadence & Step: 9.2\,s, Full raster: $\sim$36\,s & Step: 9.6\,s  \\
		Spatial sampling & 0.35$''$ & 0.35$''$ \\
		Spectral sampling & Mg\,\textsc{ii}: 50.91\,m\AA & Mg\,\textsc{ii}: 25.45\,m\AA \\
		& Si\,\textsc{iv}: 25.44\,m\AA & Si\,\textsc{iv}: 12.72\,m\AA \\
		& C\,\textsc{ii}: 25.96\,m\AA & C\,\textsc{ii}: 12.98\,m\AA \\
		\hline
	\end{tabular}
	
	\vspace{1em}
	\textbf{(c) SDO Observations} \\
	\begin{tabular}{llccc}
		\hline
		\textbf{Instrument} & \textbf{Channel} & \textbf{Wavelength} & \textbf{Cadence} & \textbf{Plate Scale} \\
		\hline
		AIA (UV) & 1600\,\AA, 1700\,\AA & UV continuum & 12\,s & 0$\arcsec$.6\,pixel$^{-1}$ \\
		AIA (EUV) & 131\,\AA, 171\,\AA, 193\,\AA, 211\,\AA, 304\,\AA, 335\,\AA & EUV emission lines & 24\,s & 0$\arcsec$.6\,pixel$^{-1}$ \\
		HMI & Continuum, LOS magnetogram & 6173\,\AA & 45\,s & 0$\arcsec$.5\,pixel$^{-1}$ \\
		\hline
	\end{tabular}
\end{table*}

\section{Observation} \label{sec2}

We analyzed two different coordinated datasets consisting of an active region (NOAA Active Region 12775), observed on 11 October 2020 from 08:04 UT to 08:35 UT (hereafter referred to as Dataset 1), and a Quiet Sun (QS) region dataset observed on 25 April 2020 from 11:08 UT to 11:42 UT (Dataset 2). The heliocentric coordinates of the target regions were ($-114\arcsec$, $-479\arcsec$) for the active region, and ($-8\arcsec$, $-144\arcsec$) for the quiet Sun regions, with corresponding observing angles of $\mu = \cos\theta = 0.86$ and $0.98$, where $\theta$ represents the heliocentric angle. The active region consisted of multiple pores and network regions with a moderate magnetic field configuration, while the quiet Sun regions were predominantly magnetically quiet. The multiwavelength overview of the chromosphere in the Dataset 1 (upper panel) and Dataset 2 (lower panel) are shown in Figure \ref{Fig1}. The field of view in all datasets is approximately $60\arcsec \times 60\arcsec$. Portions of Dataset 1 and were previously analyzed by \citet{2024ApJ...970..179C} to study the multithermal nature of spicules in the solar atmosphere.

Additionally, we analyzed a dataset obtained from the Multi Application Solar Telescope (MAST; \cite{2009A&A...501L..19M,2017CSci..113..686V}), consisting of an active region (NOAA Active Region 13599), observed on 08 March 2024 from 04:12 UT to 04:36 UT. The heliocentric coordinates of the target region were ($73\arcsec$, $-93\arcsec$), corresponding to an observing angle of $\mu = \cos\theta = 0.99$. This active region featured several diffuse sunspots and exhibited a moderate magnetic field configuration. More details about the MAST observations are provided in Section~\ref{sec2.4}.

\subsection{Swedish Solar Telescope (SST)} \label{sec2.1}

Dataset~1 includes imaging spectroscopic observations in H$\alpha$, H$\beta$, and Ca \textsc{ii} H \& K, along with imaging spectropolarimetric data in Ca \textsc{ii}~8542 \AA\ and Fe \textsc{i}~6302 \AA. Dataset~2 is similar, with imaging spectroscopy in H$\alpha$, H$\beta$, Ca \textsc{ii} H \& K, and Ca \textsc{ii}~8542 \AA, but spectropolarimetry only in Fe \textsc{i}~6302 \AA. Among these, the H$\alpha$, Ca \textsc{ii}~8542 \AA, and Fe \textsc{i}~6302 \AA\ data were acquired using the CRISP instrument \citep{scharmer2008crisp}, while the H$\beta$ and Ca \textsc{ii} H \& K observations were obtained with CHROMIS \citep{scharmer2017solarnet}. The CRISP data were originally sampled at $0\arcsec.0591$ per pixel, whereas the CHROMIS data have a finer spatial sampling of $0\arcsec.0379$ per pixel.

For Dataset~1, the average time required to scan the full spectral line was approximately 22.74s for H$\beta$ and 18.98s for Ca \textsc{ii} H \& K. However, this cadence was not uniform throughout the time series. In several instances, the time interval between successive scans increased to two or even three times the nominal cadence (17.59s for H$\beta$ and 17.31s for Ca \textsc{ii} H \& K) due to occasional missing frames. To address these irregularities and ensure uniformly sampled time series, linear interpolation was applied to fill the missing frames. This step is necessary for the wavelet analysis of the time-varying signal (see Section \ref{ssec3.4}), which requires data sampled at constant time intervals. As a result, the H$\beta$ and Ca \textsc{ii} H \& K time series were resampled to uniform cadences of 17.59s and 17.31s, respectively.

A similar approach was followed for Dataset 2, where the average time cadence of the H$\beta$ observations was 25.16\,s, while the Ca\,\textsc{ii} H \& K cadence was nearly constant at approximately 16.62\,s. Due to the presence of missing frames in the H$\beta$ time series, linear interpolation was again performed to obtain a uniform cadence of 16.59\,s.

To ensure spatial consistency across all datasets, the CRISP data were bilinearly interpolated to match the finer CHROMIS pixel scale. Further details about the SST observations and their characteristics for both datasets are summarized in the upper panel (a) of Table~\ref{tab1}.

\subsection{Interface Region Imaging Spectrograph (IRIS)} \label{sec2.2}

The two coordinated IRIS\footnote{IRIS OBSIDs: 3630108417 and 3620008803 (Dataset~1 and Dataset~2)} observations, including both spectral data and Slit-Jaw Images (SJIs), were retrieved from the IRIS data archive\footnote{\url{https://iris.lmsal.com/search/}}. The spectral data were the wavelength-calibrated Level 2 observations, which use neutral lines for absolute wavelength calibration \citep{2014SoPh..289.2733D} and are assumed to form in the lower atmosphere exhibiting intrinsic velocities below 1 km s$^{-1}$. Detailed information about these IRIS observations is provided in the middle panel (b) of Table\ref{tab1}.

For Dataset~1, it is important to note that while the average temporal cadence of the Si \textsc{iv}~1400 \AA\ observations is 21.88s, the cadence between successive images is not uniform. Specifically, every sixth image in the Si \textsc{iv}~1400 \AA\ sequence is captured at approximately double the interval of the preceding five images, which have a cadence of about 18.22s. To achieve a uniform cadence across the dataset, linear interpolation was applied to every sixth image, effectively resampling the entire Si \textsc{iv}~1400 \AA\ time series to a constant cadence of 18.22s.

In contrast, for Dataset 2, the cadence of all SJIs was nearly uniform, with average values of 28.68s. To ensure accurate spatial alignment between the SST and IRIS-SJI images, the SJIs were first bilinearly interpolated to match the CHROMIS plate scale. Subsequently, a cross-correlation was performed between the nearly simultaneous IRIS 2832 \AA\ SJI image and the photospheric wing image from SST H$\alpha$.

\subsection{Solar Dynamics Observatory (SDO)} \label{sec2.3}

The corresponding registered coordinated cutouts were downloaded using the SDO cutout service, which includes EUV and UV data from SDO/AIA and continuum images and magnetograms from SDO/HMI. Detailed information about the SDO observations can be found in lower panel (c) of Table~\ref{tab1}. To achieve precise alignment between the SDO images and the SST observations, the individual AIA channels were first coaligned with each other. These coaligned AIA images were then bilinearly interpolated to match the SST plate scale. Following this, a cross-correlation was performed between the nearly simultaneous AIA 1700~\AA{} and the photospheric wing image from SST H$\alpha$ to establish the coordinated alignment of SDO/AIA with the SST observations. A similar procedure was applied to the HMI continuum images to get coaligned HMI continuum and magnetogram data.

\subsection{Multi Application Solar Telescope (MAST)}  \label{sec2.4}

We use only Ca\,\textsc{ii} 8542~\AA\ line scan observations to study few EBs in the chromosphere. The Ca\,\textsc{ii} 8542~\AA\ imaging spectroscopic observations were conducted using the narrowband imager of the 50\,cm off-axis Gregorian MAST at Udaipur, India, under moderate seeing conditions. The narrowband imager (\cite{2014SoPh..289.4007R,2017SoPh..292..106M,2017SoPh..292...49T,2024MNRAS.535.1228D}), equipped with two voltage-tunable lithium niobate Fabry–Pérot etalons, enables two-dimensional spectral imaging. Observations were performed in the Ca\,\textsc{ii} 8542~\AA\ line with a unequal spectral sampling over a range of $\pm$0.75~\AA\ from the line center, acquiring three filtergrams per spectral point with an exposure of 0.7\,s each. Prior to the scan, the line center was determined to ensure symmetric spectral coverage, and calibration frames (15 darks, 200 flats) were recorded for flat-fielding and dark-correction. Spectral tuning was achieved via voltage control, with field-dependent wavelength shifts corrected using a reference scan through a diffuser.

\begin{figure}
	\centering
	\includegraphics[width=85mm]{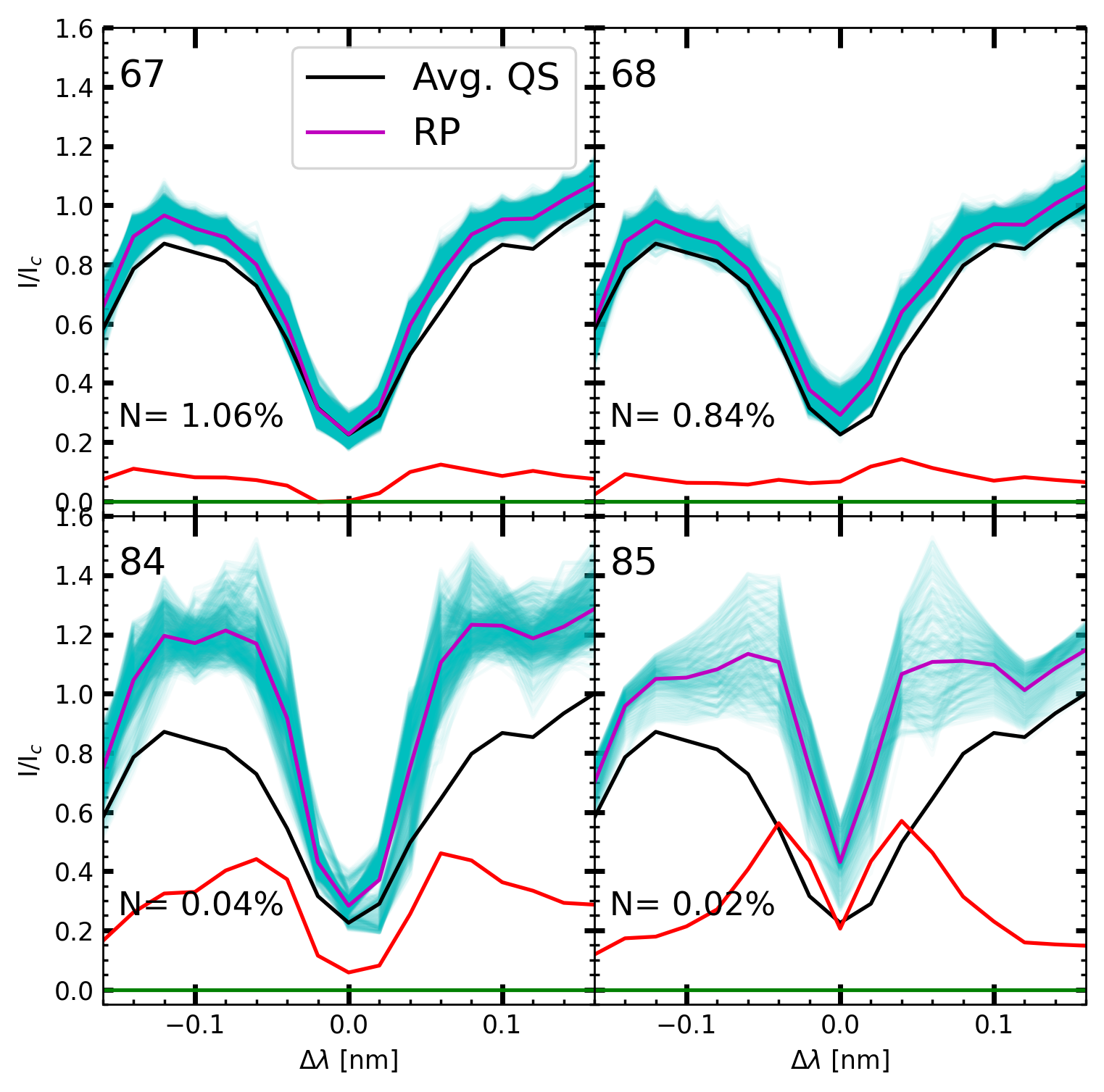}	
	\caption{Four of the 85 representative profiles (RPs) identified via \textit{k}-means clustering of an H$\beta$ line-scan dataset.  
		Magenta curves in the upper panels (RPs~67 and 68) show cluster centers of Magnetic Concentrations, while those in the lower panels (RPs~84 and 85) correspond to EBs. Cyan shading marks all profiles in each cluster. Red curves show the difference between the cluster center and the average profile (black), with the green line indicating zero normalized intensity. $N$ denotes the percentage of total profiles in each cluster.}
	
	\label{Fig2}
\end{figure}

\begin{figure*}
	\centering
	\includegraphics[width=85mm]{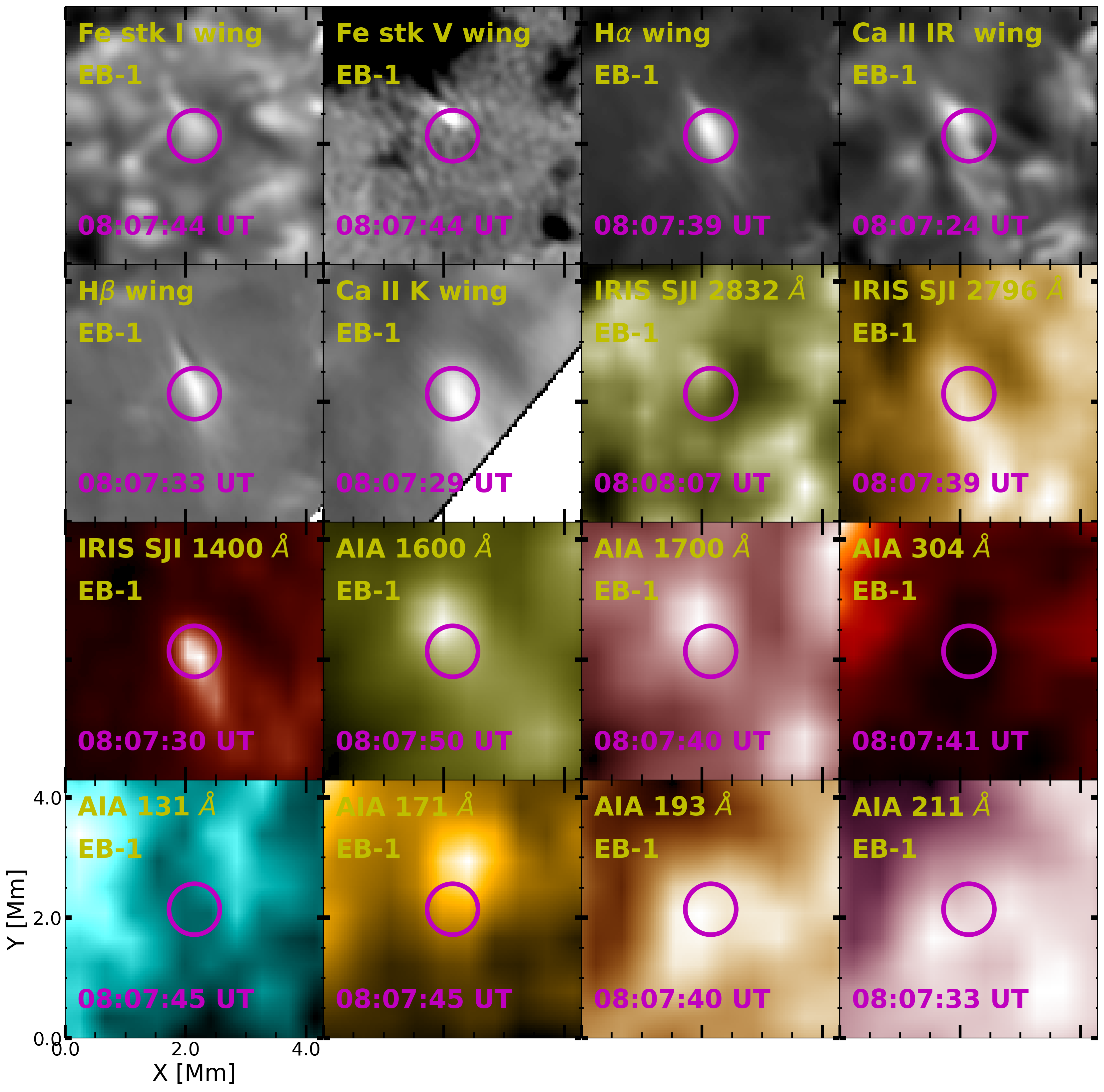}
	\includegraphics[width=85mm]{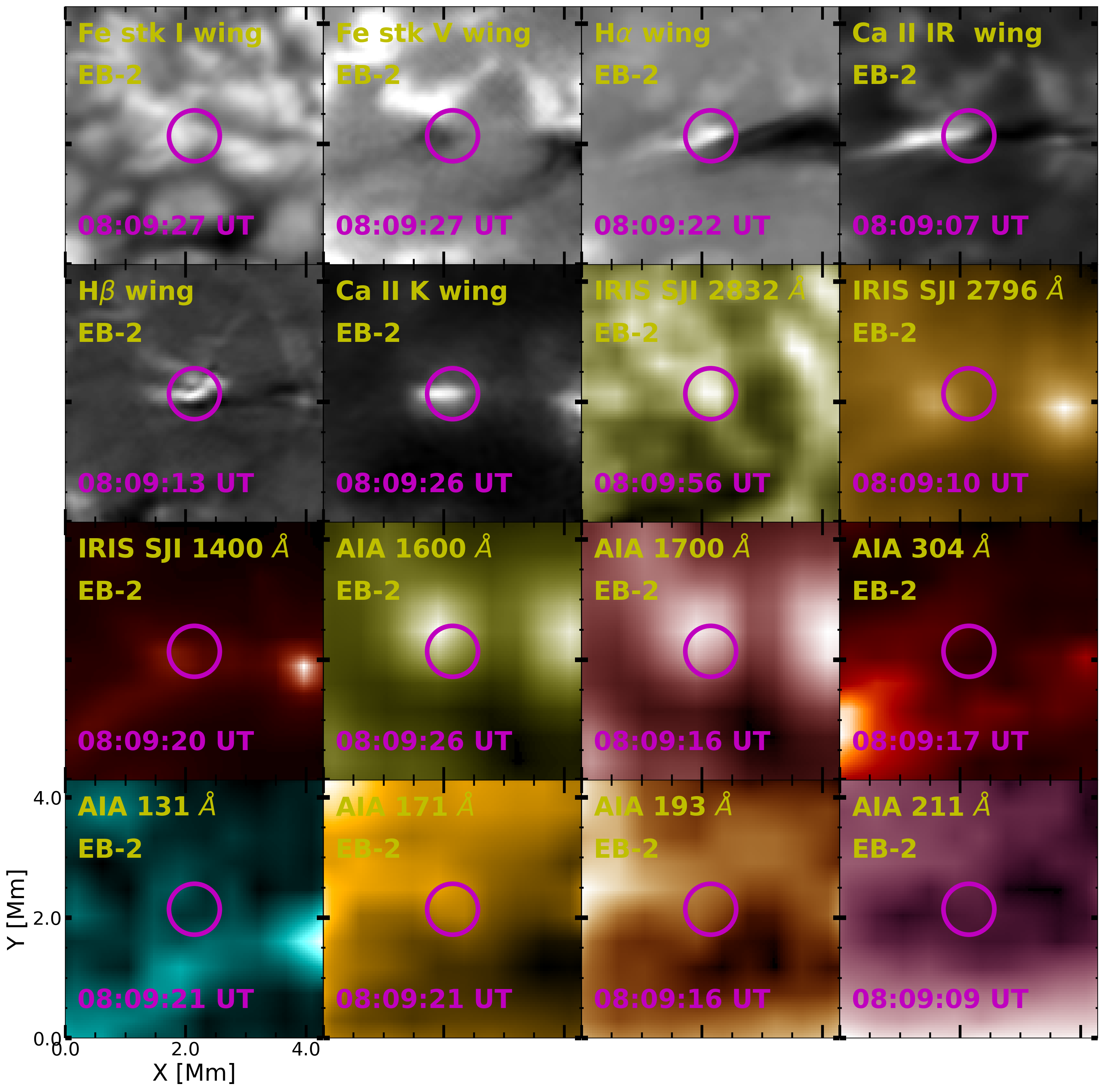}
	\includegraphics[width=85mm]{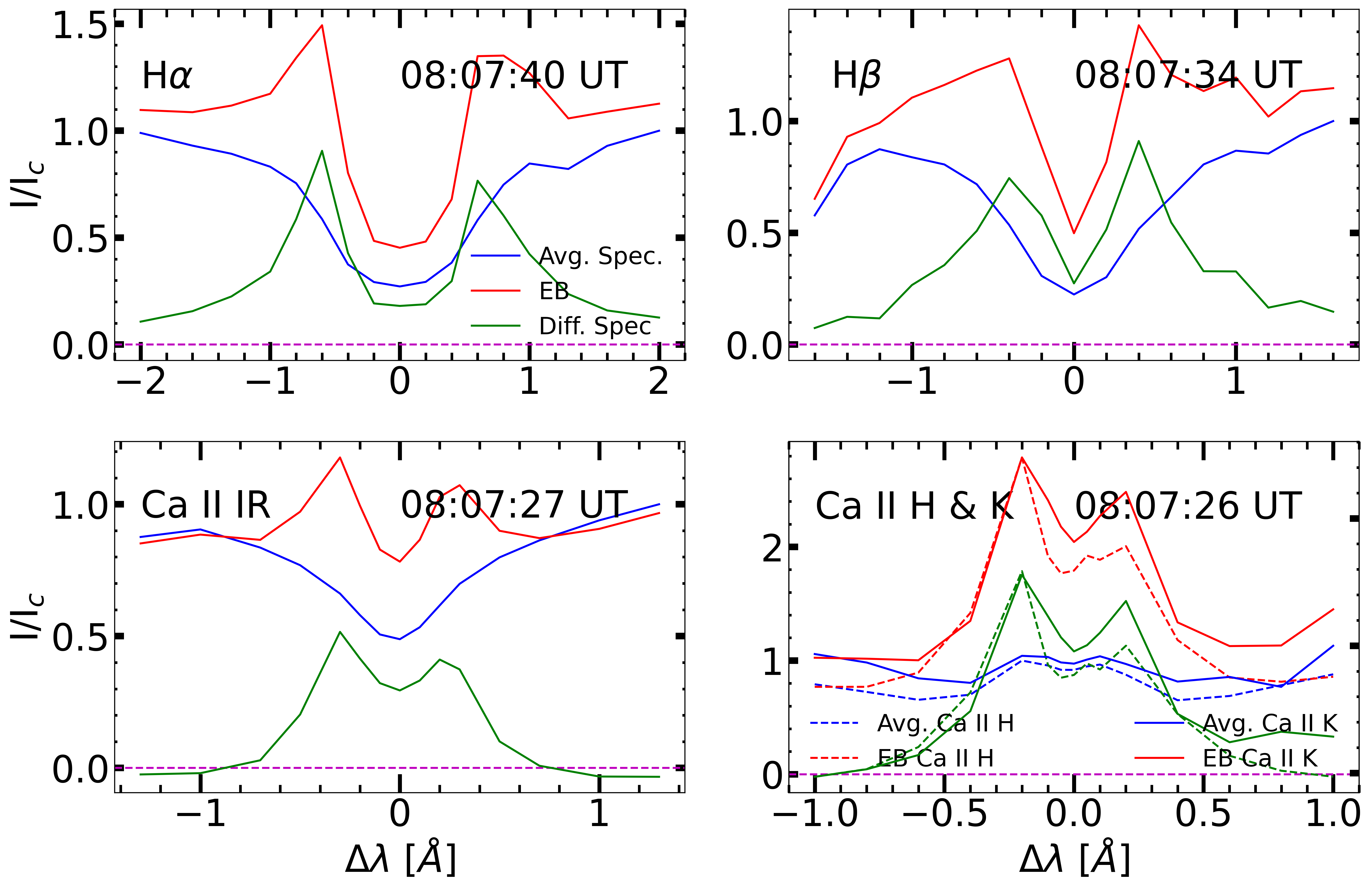}
	\includegraphics[width=85mm]{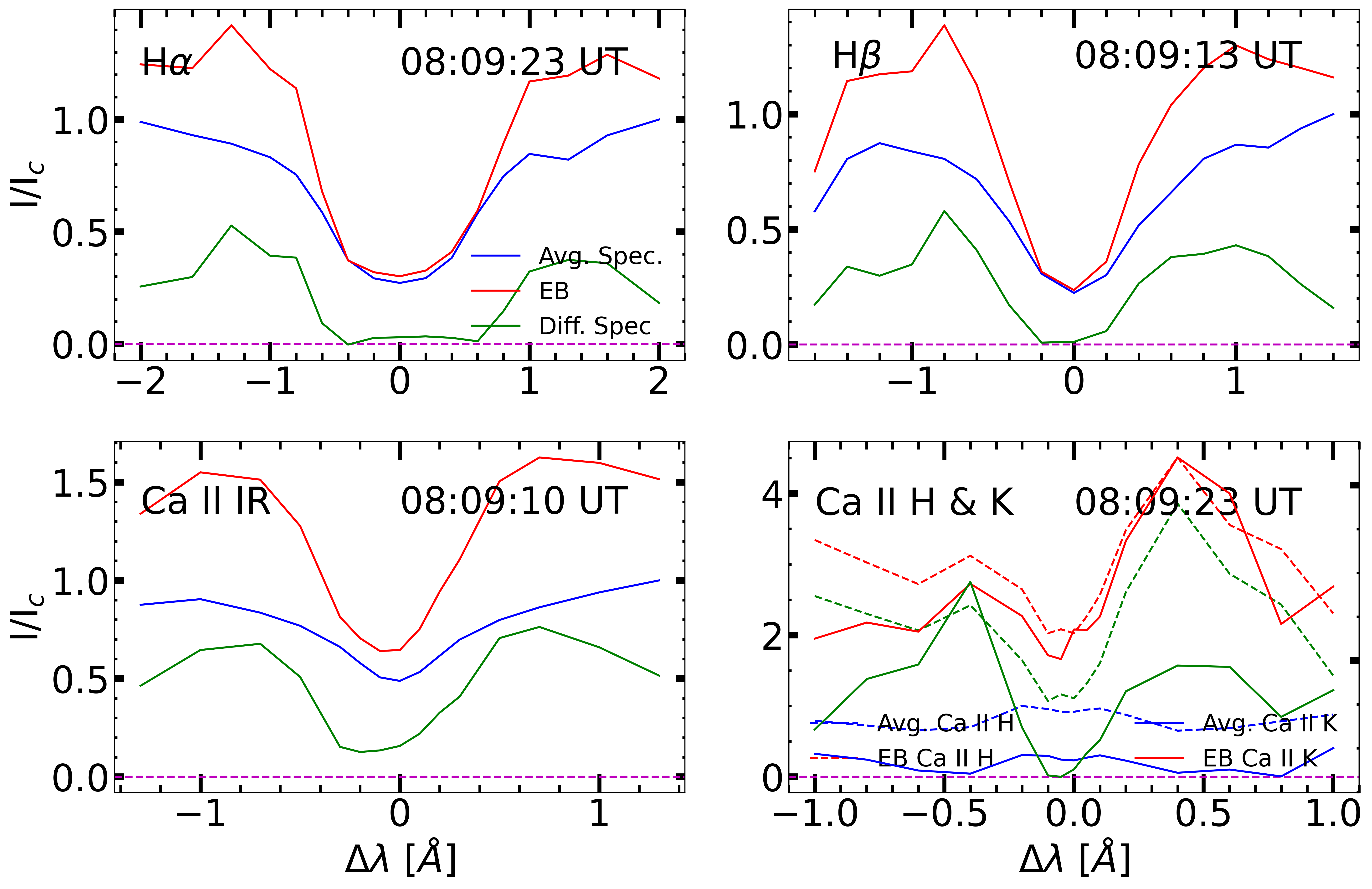}
	
	\caption{
		\textbf{Top panel:} Coordinated multi-wavelength observations of two EBs (labeled as colored 1 and 2 in the upper panel of Figure \ref{Fig1}) spanning from the lower to upper solar atmosphere, including Fe I 6302~\AA\ wing, Stoke V of Fe I 6302~\AA, H$\alpha$ wing, Ca\,\textsc{ii} 8542~\AA\ wing, H$\beta$ wing, Ca\,\textsc{ii} K wing, IRIS SJI 2832~\AA, 2796~\AA, 1400~\AA, and SDO/AIA 1600~\AA, 1700~\AA, 304~\AA, 131~\AA, 171~\AA, 193~\AA, and 211~\AA\ channels. 
		\textbf{Bottom panel:} Spectral profiles at the EB locations in H$\alpha$, H$\beta$, Ca\,\textsc{ii} 8542~\AA, and Ca\,\textsc{ii} H \& K. Red and blue lines represent EB and quiet Sun spectra, respectively, while green and magenta lines show the difference (EB minus quiet Sun) and the normalized zero level. The spectra exhibit typical wing enhancements characteristic of EBs. Although EBs are observed nearly co-spatially and co-temporally in chromospheric wings, certain IRIS SJIs, and AIA 1600/1700~\AA\ channels, they remain undetected in hotter AIA channels. The animation spans a total duration of $\sim$ 2 seconds and consists of 29 frames for EB-1 and 19 frames for EB-2, illustrating the temporal evolution of the EBs in the H$\beta$ spectral line. The animation of this Figure is available online.}
	\label{Fig4}
\end{figure*}

\section{Results \& Discussion}\label{sec3}

Building upon previous studies of EBs and QSEBs in the solar atmosphere, we identify these phenomena as exhibiting distinct dynamics compared to other chromospheric features. This distinction is based on their characteristic signature: a significant enhancement in the wings of chromospheric spectral lines with respect to the quiet Sun spectra. To classify EBs and QSEBs in Dataset 1 and Dataset 2, we employ \textit{k}-means clustering \citep{Everitt_1972} on the H$\beta$ observations (see Section~\ref{ssec3.1}). This technique allows us to isolate pixels corresponding to EBs and QSEBs. Based on this classification, we manually select several representative events to further investigate their impact and signatures across different temperature regimes in the solar atmosphere. The analysis and results from these investigations are presented in Section~\ref{ssec3.2}. Additionally, besides the SST Ca\,\textsc{ii}~8542~\AA\ observations, we also examine a few EB events using Ca\,\textsc{ii}~8542~\AA\ data obtained from the MAST. The complementary examples of these EBs are presented in Section~\ref{ssec3.2}.  

To explore the vertical structuring of these events in the chromosphere, we employed the STockholm inversion Code\footnote{\url{https://github.com/jaimedelacruz/stic}}
(STiC; \cite{2016ApJ...830L..30D,2019A&A...623A..74D}) to retrieve the stratification of physical parameters in the lower atmosphere (See Section~\ref{ssec3.3} for more details about STiC). This inversion allowed us to examine how the thermal and physical properties evolve with optical depth during EBs and QSEBs events. 

Moreover, we explore the rate of episodic reconnection of few EBs in the lower chromosphere. The relevant results are presented in Section~\ref{ssec3.4}. Finally, we investigate the possible connection between the occurrence of EBs/QSEBs and the origin of spicules. This analysis provides insight into whether EBs contribute to the formation or triggering of spicular structures. The results of this investigation are also discussed in Section~\ref{ssec3.5}.

\begin{figure*}
	\centering
	\includegraphics[width=85mm]{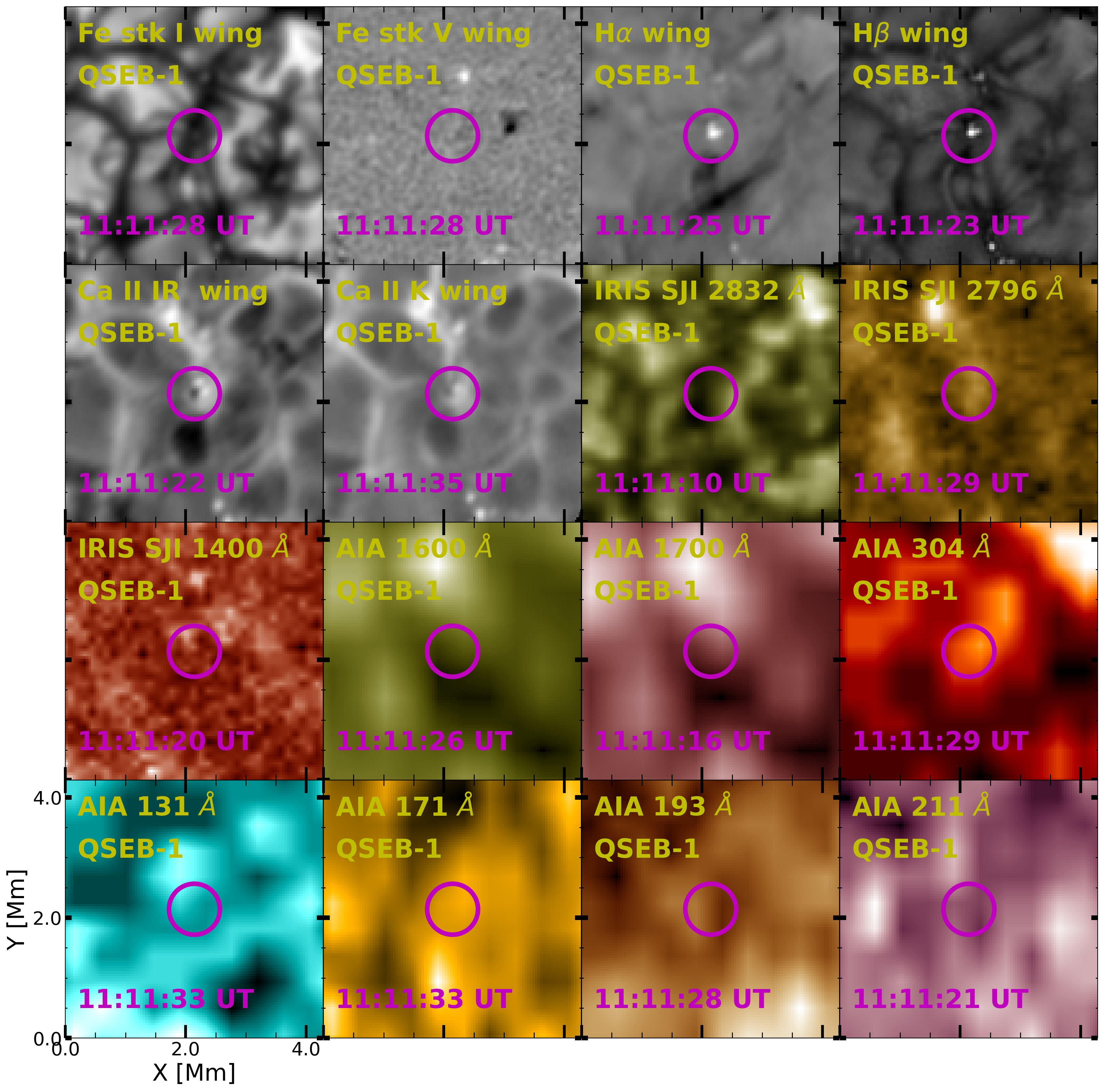}
	\includegraphics[width=85mm]{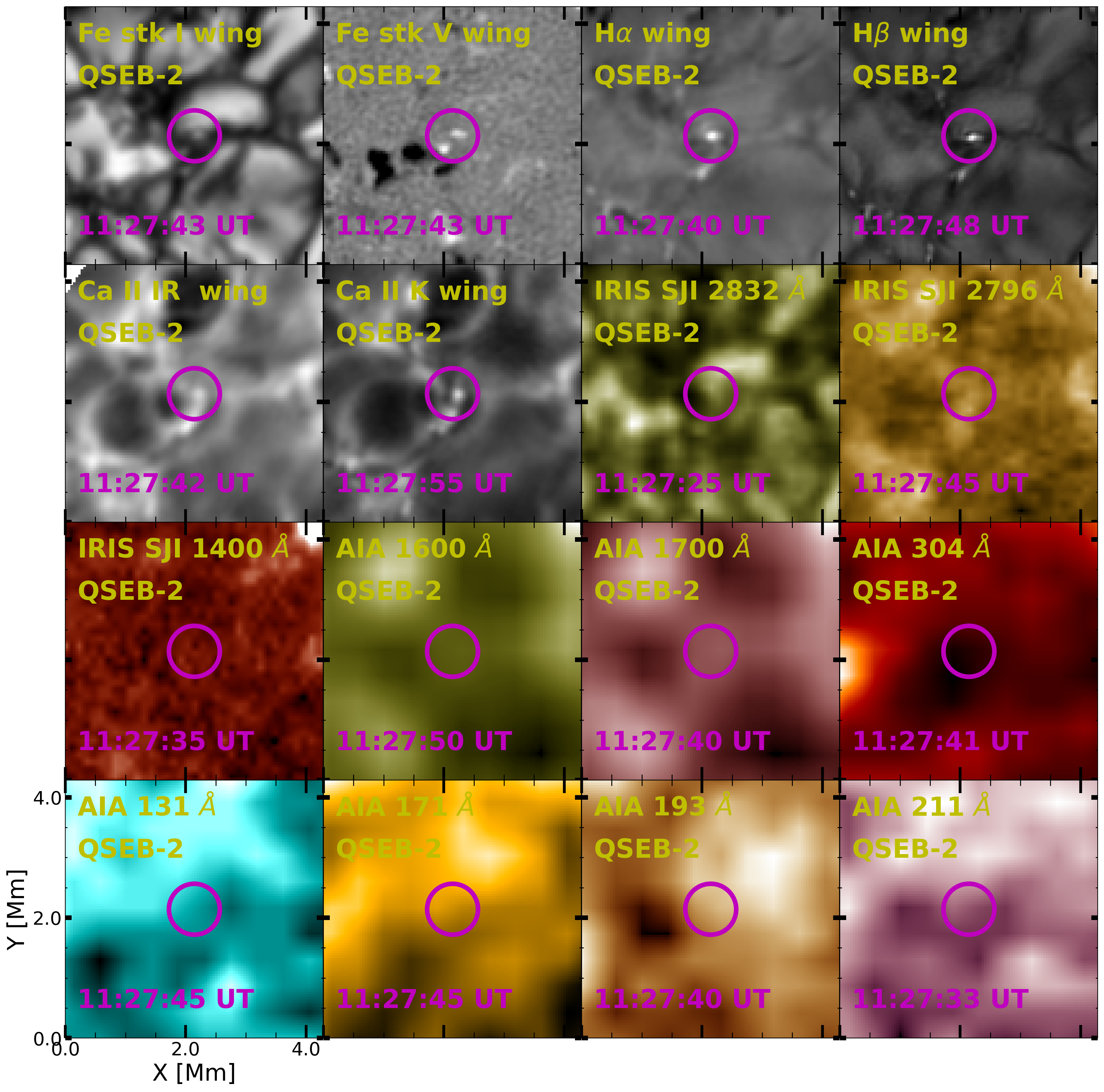}
	\includegraphics[width=85mm]{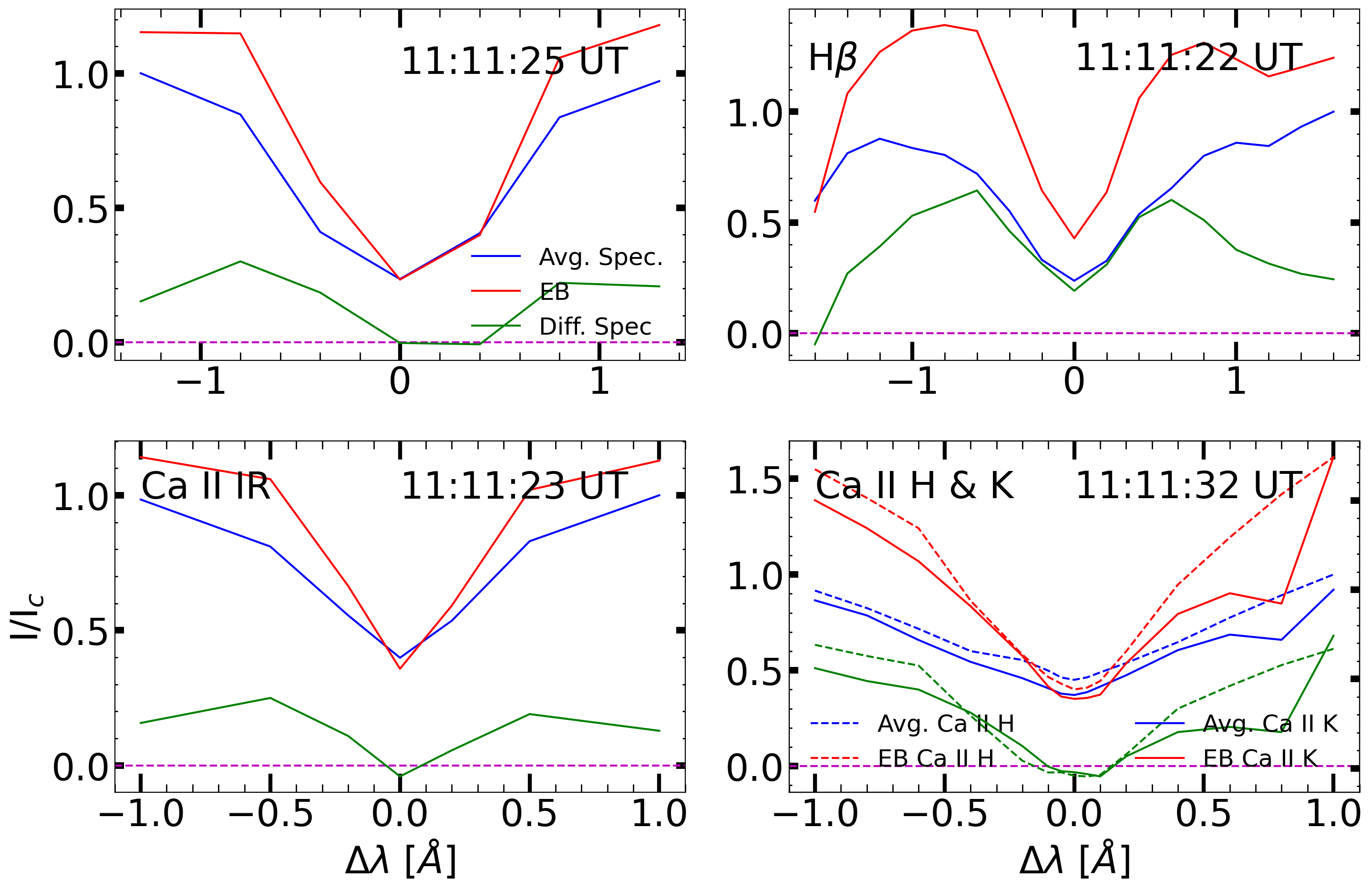}
	\includegraphics[width=85mm]{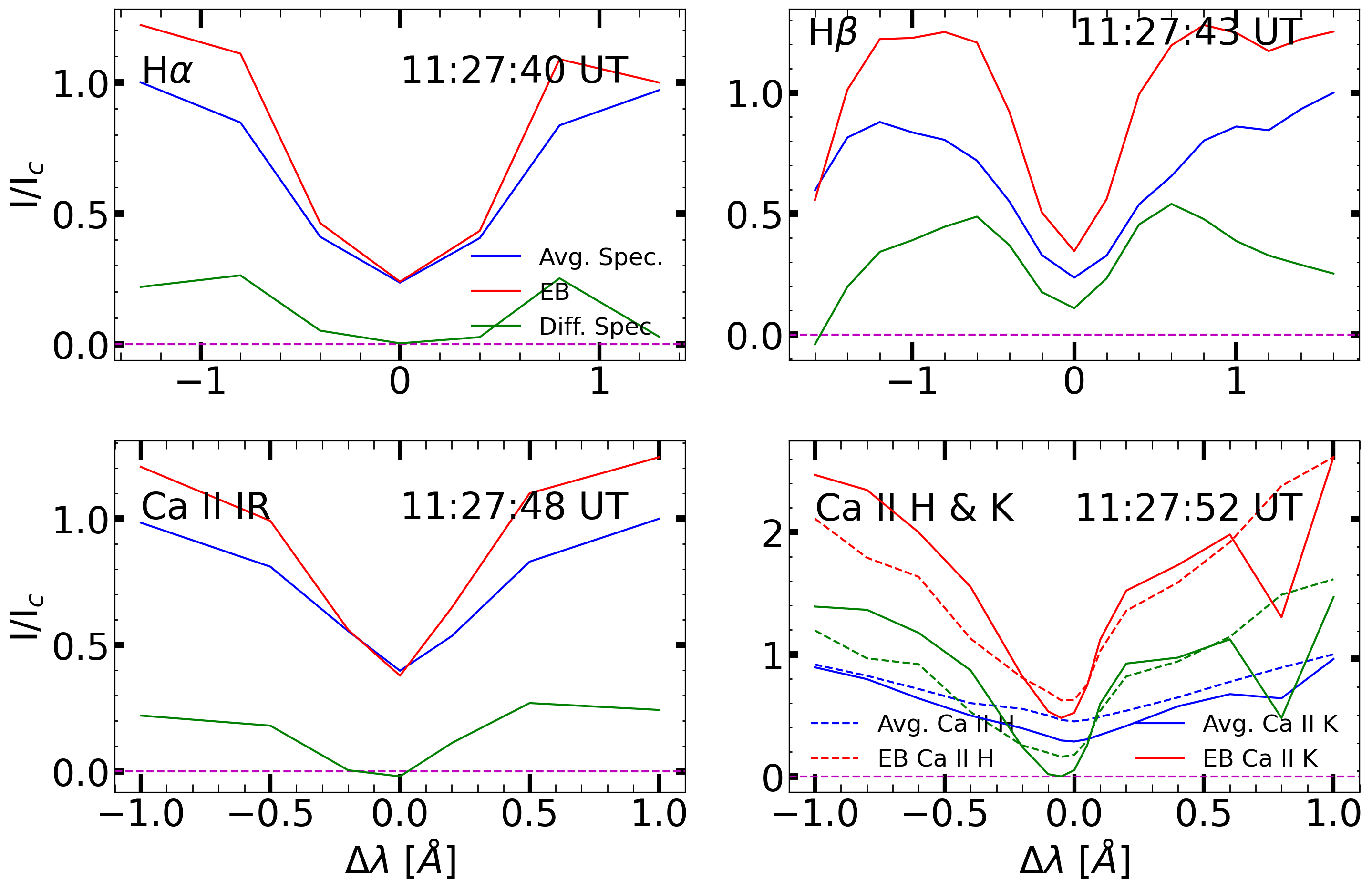}
	
	\caption{\textbf{Top panel:} Coordinated multi-wavelength observations of two QSEBs (labeled as colored 1 and 2 in the lower panel of Figure \ref{Fig1}) spanning from the lower to upper solar atmosphere, including Fe I 6302~\AA\ wing, Stoke V of Fe I 6302~\AA, H$\alpha$ wing, Ca\,\textsc{ii} 8542~\AA\ wing, H$\beta$ wing, Ca\,\textsc{ii} K wing, IRIS SJI 2832~\AA, 2796~\AA, 1400~\AA, and SDO/AIA 1600~\AA, 1700~\AA, 304~\AA, 131~\AA, 171~\AA, 193~\AA, and 211~\AA\ channels. \textbf{Bottom panel:} Spectral profiles at the QSEB locations in H$\alpha$, H$\beta$,Ca\,\textsc{ii} 8542~\AA, and Ca\,\textsc{ii} H \& K. Red and blue lines represent EB and quiet Sun spectra, respectively, while green and magenta lines show the difference (QSEB minus quiet Sun) and the normalized zero level. The spectra exhibit typical wing enhancements characteristic of QSEBs. Although QSEBs are observed nearly co-spatially and co-temporally in chromospheric wings and IRIS SJIs, and AIA 1600/1700~\AA\ channels, they remain undetected in hotter AIA channels. The animation spans a total duration of $\sim$ 1 second and consists of 9 frames for QSEB-1 and 15 frames for QSEB-2, illustrating the temporal evolution of the QSEBs in the H$\beta$ spectral line. The animation of this Figure is available online.}
	\label{Fig6}
\end{figure*}

\begin{figure}
	\centering
	\includegraphics[width=50mm]{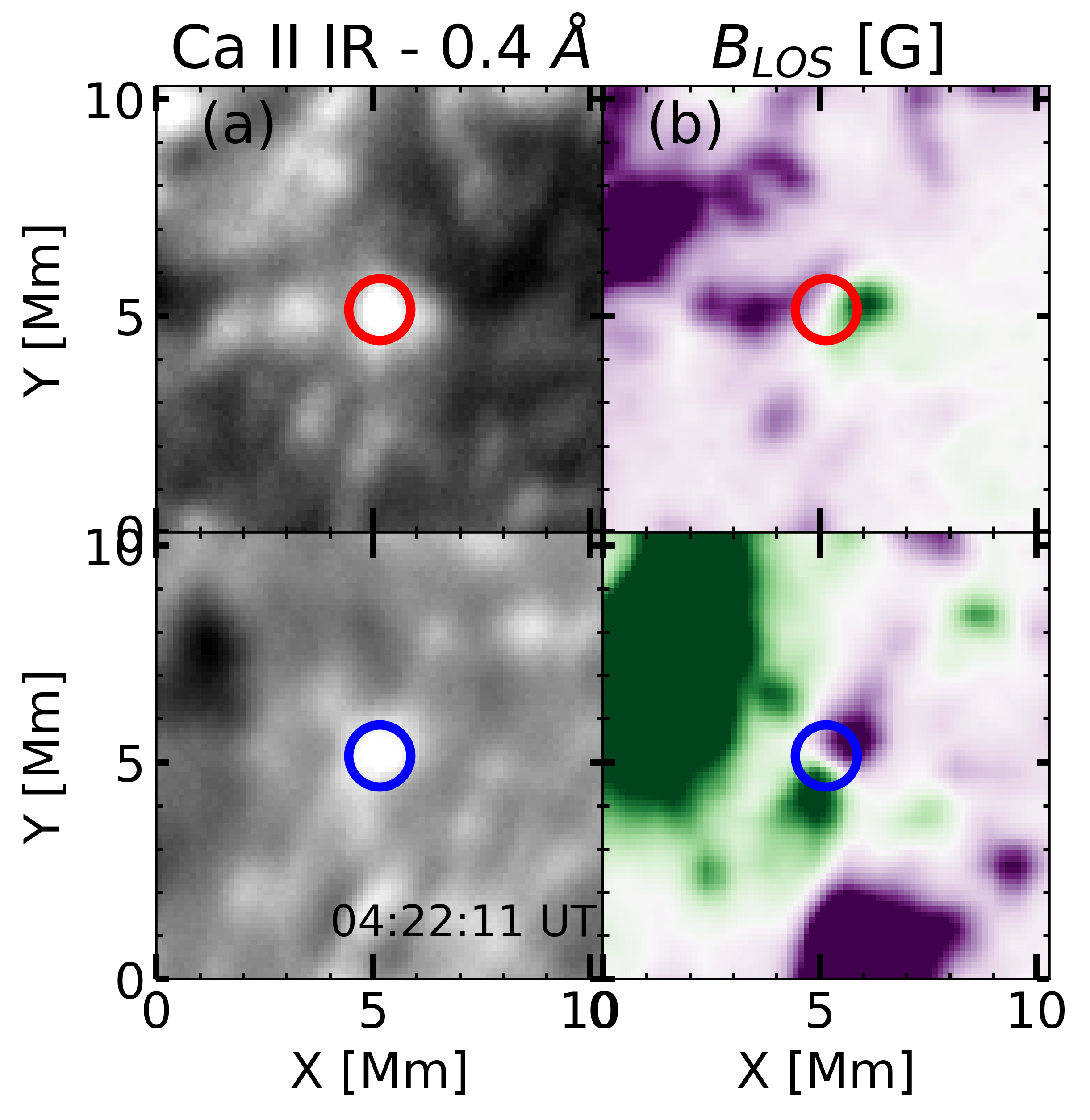}
	\includegraphics[width=29mm]{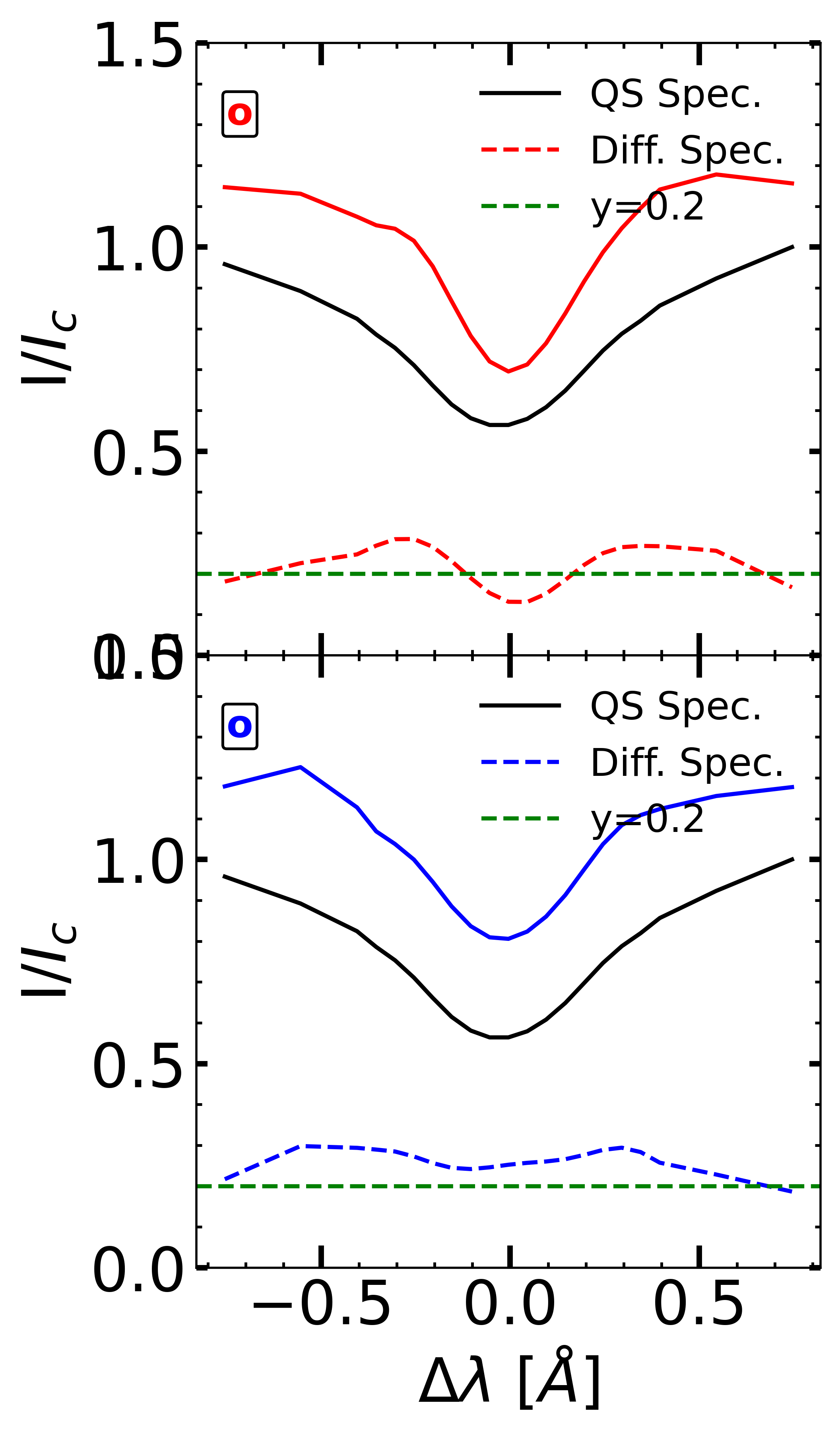}
	\caption{Complementary examples of EBs observed with MAST in the Ca\,\textsc{ii} 8542~\AA\ line. Panels (a) and (b) show the low chromosphere as seen in the Ca\,\textsc{ii} 8542~\AA\ wing and the corresponding HMI magnetogram respectively. The right panel presents the spectral profiles of the EBs and the QS respectively.}
	\label{Fig5}
\end{figure}

\begin{figure}
	\centering
	\includegraphics[width=85mm]{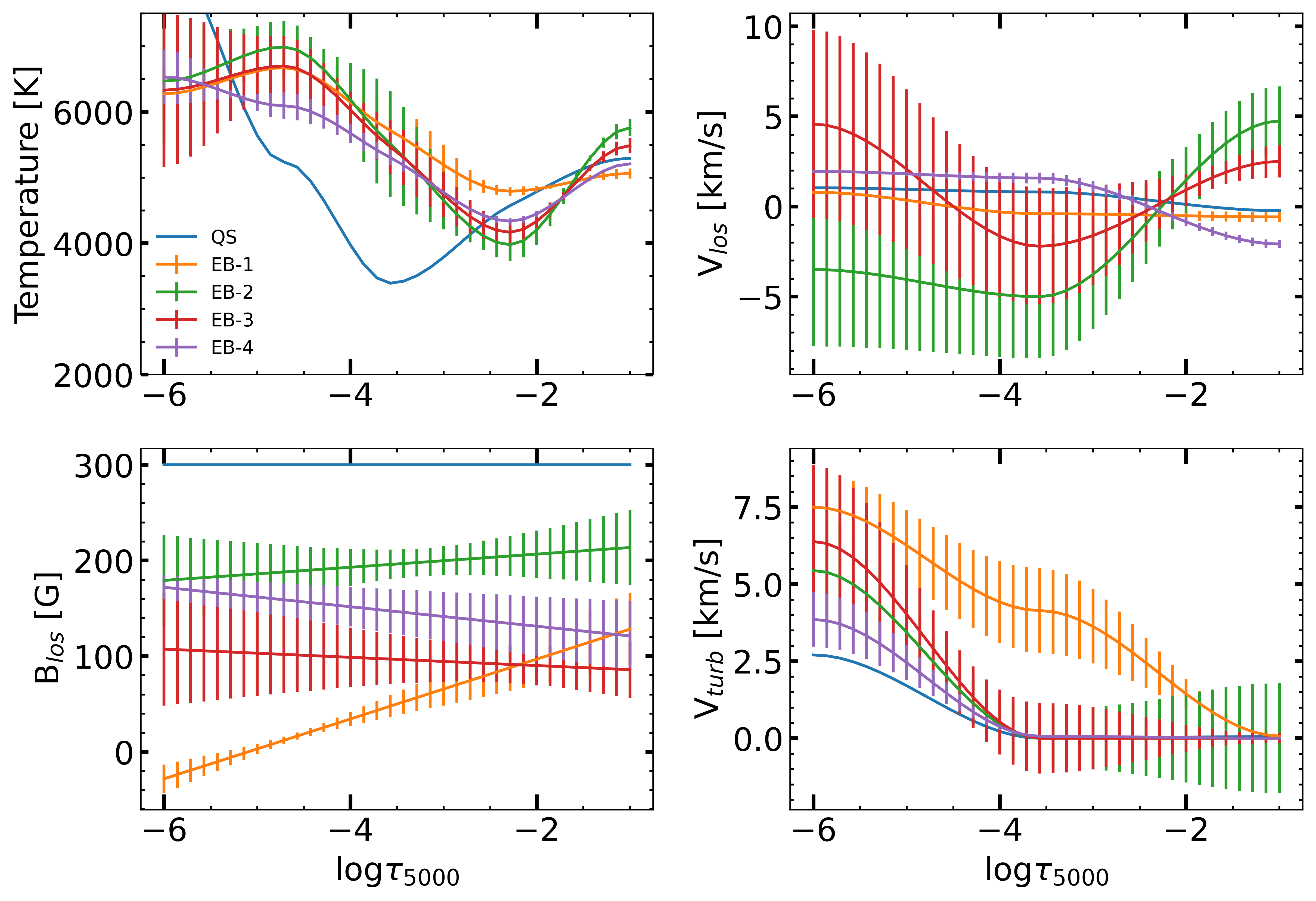}	
	\caption{The upper panels show the comparison of Temperature and $V_{\rm LOS}$, whereas the lower panels represent the $B_{\rm LOS}$ and $V_{\rm turb}$ stratification as a function of optical depth for four EBs, labeled as 1, 2, 3, and 4 in Figure~\ref{Fig1}, along with the QS reference. These inferred parameters are obtained from STiC inversions of the full Stokes profiles of the Ca\,\textsc{ii} 8542~\AA\ spectral line. The vertical error bars represent the 1$\sigma$ uncertainties of the derived parameters, estimated using a Monte Carlo approach by perturbing the model atmosphere and repeating the inversions 500 times.} The enhanced magnitude in $T$ with respect to QS profile indicates localized heating in the lower chromosphere associated with the occurrence of the EBs.
	\label{Fig_STiC_EBs}
\end{figure}

\begin{figure}
	\centering
	\includegraphics[width=85mm]{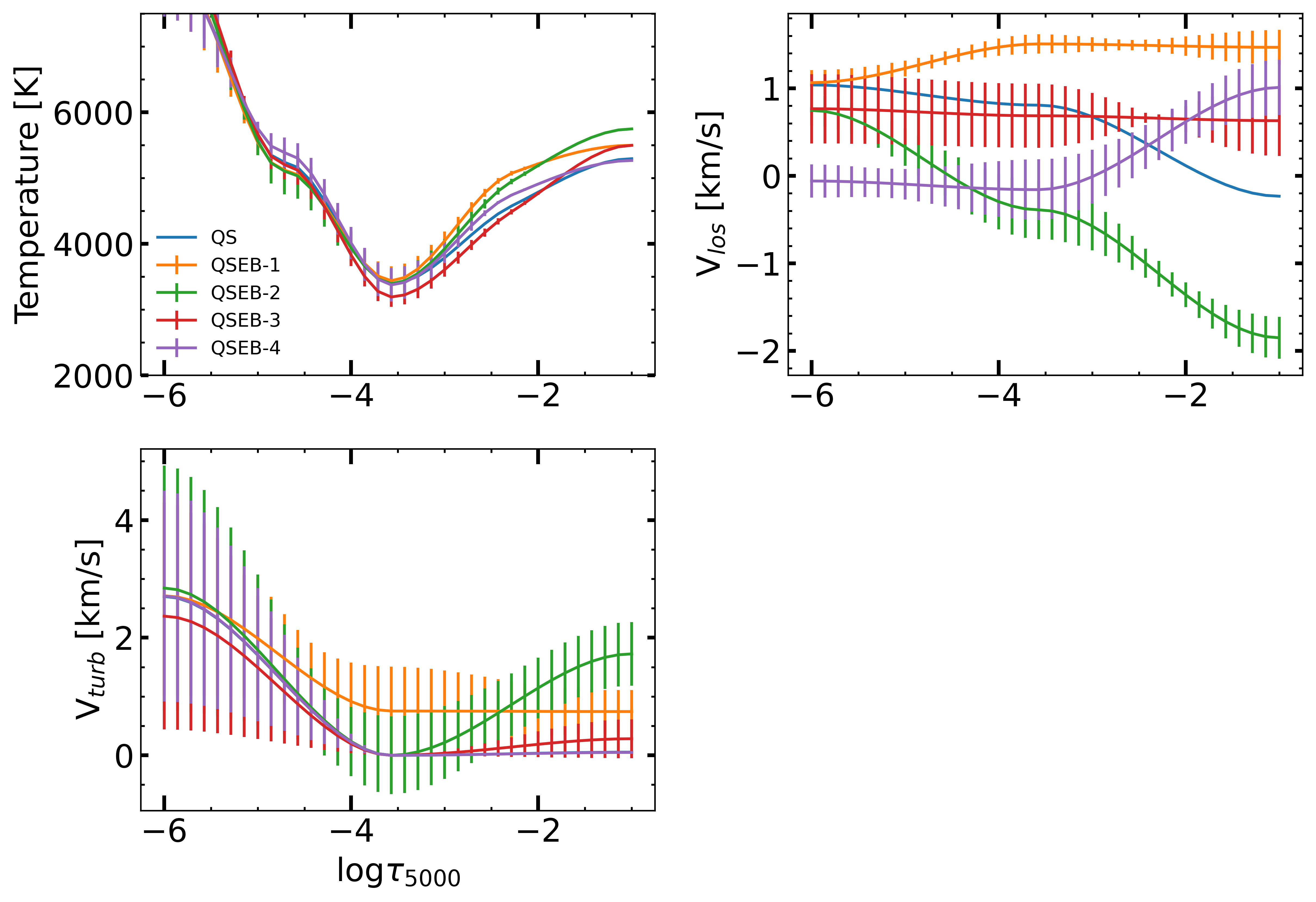}	
	\caption{Similar to Figure \ref{Fig_STiC_EBs} but for the four QSEBs, labeled as 1, 2, 3, and 4 in the lower panel of Figure~\ref{Fig1}, along with the QS reference.}
	\label{Fig_STiC_QSEBs}
\end{figure}

\subsection{Identification of EBs and QSEBs using \textit{k}-means clustering on H$\beta$ spectral line}
\label{ssec3.1}

We employ the \textit{k}-means clustering technique (implemented in Scikit-learn; \citep{JMLR:v12:pedregosa11a}) on H$\beta$ imaging spectroscopy data (Dataset 1 and Dataset 2) to classify pixels based on their spectral characteristics. This \textit{k}-means clustering was applied directly to all normalized H$\beta$ spectral profiles within the field of view, without prior correction for Doppler shifts, in order to preserve spectral asymmetries and Doppler-shifted wing enhancements. Clusters corresponding to EBs/QSEBs were identified by examining the cluster centers (representative profiles) and selecting those exhibiting characteristic wing enhancements relative to the average quiet-Sun profile. The associated spectra within these clusters were further verified to show similar wing-enhanced behavior. Although the elbow method was used to estimate the optimal number of clusters, lower cluster numbers (40–-80) resulted in incomplete segregation of EB-like profiles with different spectral morphologies. Increasing the number of clusters improved the separation of distinct spectral classes and yielded more homogeneous cluster centers, leading to the adoption of 85 clusters.

Compared with intensity-threshold methods, \textit{k}-means clustering utilizes the entire spectral profile rather than intensity enhancements at selected wavelength positions. Since EBs/QSEBs often exhibit variable and asymmetric wing enhancements, threshold-based approaches may miss or underestimate events whose strongest emission occurs at different wavelength offsets or preferentially in one wing. By considering the overall spectral morphology, \textit{k}-means clustering is better suited to capture such variations.

It should be noted that we utilize H$\beta$ imaging spectroscopic observations, as it offer improved spatial resolution and enhanced intensity contrast compared to H$\alpha$ observations. The shorter wavelength of the H$\beta$ line contributes to higher spatial resolution, while the increased contrast arises from the wavelength dependence of the Planck function, which provides greater sensitivity to temperature enhancements at shorter wavelengths such as those occurring at magnetic reconnection sites. These advantages enable more precise identification of various chromospheric features in the solar atmosphere.

Figure~\ref{Fig2} presents only four representative clusters (two each corresponding to Magnetic Concentrations (MCs), and EBs) out of the 85 identified, in order to facilitate a clear comparison between MCs and EBs. Such a distinction is essential, as previous studies have warned that MCs may also appear bright and can therefore be misidentified as EBs \citep{1917ApJ....46..298E,2015ApJ...812...11V}. A similar \textit{k}-means clustering approach was applied to segregate pixels corresponding to QSEBs in Dataset~2 but those results are not included here for brevity.

It should be noted that apart from the H$\beta$ line, the Ca\,\textsc{ii}~8542~\AA\ line, despite its longer wavelength and therefore slightly lower spatial resolution, also present a powerful diagnostic for studying EBs and QSEBs. This is because these events predominantly occur in the lower chromosphere, where the Ca\,\textsc{ii}~8542~\AA\ line is primarily formed. In addition, this line is highly sensitive to temperature and velocity perturbations, making it effective for tracing the dynamic and thermal evolution of such small-scale reconnection events (\cite{2012A&A...543A..34D,2016ApJ...826L..10S}).

\subsection{Multi wavelength signature of EBs and QSEBs  in the solar atmosphere} \label{ssec3.2}

Following the detection of EBs and QSEBs using \textit{k}-means clustering, we selected four individual examples of both EBs and QSEBs to study their signatures across multiple wavelengths corresponding to different temperature regimes in the solar atmosphere. Figures \ref{Fig4} and \ref{Fig6} present observations of two EBs (labeled as colored 1 and 2 in upper panel of Figure \ref{Fig1}) and two QSEBs (also labeled as colored 1 and 2 in the lower panel of Figure \ref{Fig1}), respectively, across wavelengths spanning from the low to the upper solar atmosphere. Two additional examples of EBs (labeled as colored 3 and 4 in the upper panel of Figure \ref{Fig1}) and QSEBs (also labeled as colored 3 and 4 in lower panel of Figure \ref{Fig1}) are shown in Figure \ref{FigA1} and Figure \ref{FigA3} in Appendix \ref{Apd1}, respectively, in the same format. Additionally, we present four complementary examples of EBs observed in Ca\,\textsc{ii} 8542 \AA~ from MAST: two examples are shown in Figure \ref{Fig5}, and the remaining two are included in Figure \ref{FigA2} in Appendix \ref{Apd1}.

To further investigate the response of IRIS transition-region spectral lines to EBs, we selected events that were simultaneously observed by both the IRIS slit and the SST. Owing to the limited spectral field of view of IRIS in Dataset~1, only two EB events (labeled 5 and 6 in Figure~\ref{Fig1}) coincided with the IRIS slit. In contrast, the IRIS observations corresponding to Dataset~2 were obtained in sit-and-stare mode and did not capture any clear QSEBs within the slit. Consequently, the transition-region temperature response of QSEBs could not be examined in this study.

Our results suggest that both EBs and QSEBs typically show strong enhancements in both wings of chromospheric spectral lines such as H$\alpha$, H$\beta$, Ca \textsc{ii}~8542~\AA, and Ca \textsc{ii} H \& K (though one wing may dominate in intensity). This characteristic wing enhancement is a well-established signature of EBs (as reported in earlier studies e.g., \citealt{2002ApJ...575..506G,Pariat.et.al.07,2013ApJ...779..125N}) and QSEBs which exhibit a similar spectral behavior (\cite{2016A&A...592A.100R,2020A&A...641L...5J,2022A&A...664A..72J}). While EBs exhibit opposite magnetic polarity in the photosphere, consistent with previous findings (\citealt{2011ApJ...736...71W,2013JPhCS.440a2007R,2013ApJ...779..125N,2015ApJ...812...11V}), QSEBs lack clear opposite polarities signatures. This is consistent with the results of \cite{2020A&A...641L...5J}, who suggested that the absence of detectable polarity inversion in QSEBs may arise from limited spatial resolution or intrinsically weaker magnetic fields. However, the absence of opposite polarity does not necessarily exclude magnetic reconnection, as \cite{2002ApJ...575..506G} have shown that EBs may also be triggered in unipolar magnetic configurations through the interaction of topologically different magnetic structures. Furthermore, recent studies of QSEBs have reported fan-spine magnetic topologies associated with these events \citep{2025A&A...698A.174B,2025A&A...693A.221B}, suggesting that reconnection may occur in complex three-dimensional magnetic configurations in addition to simple bipolar flux-cancellation scenarios.

\begin{figure*}
	\centering
	\includegraphics[width=85mm]{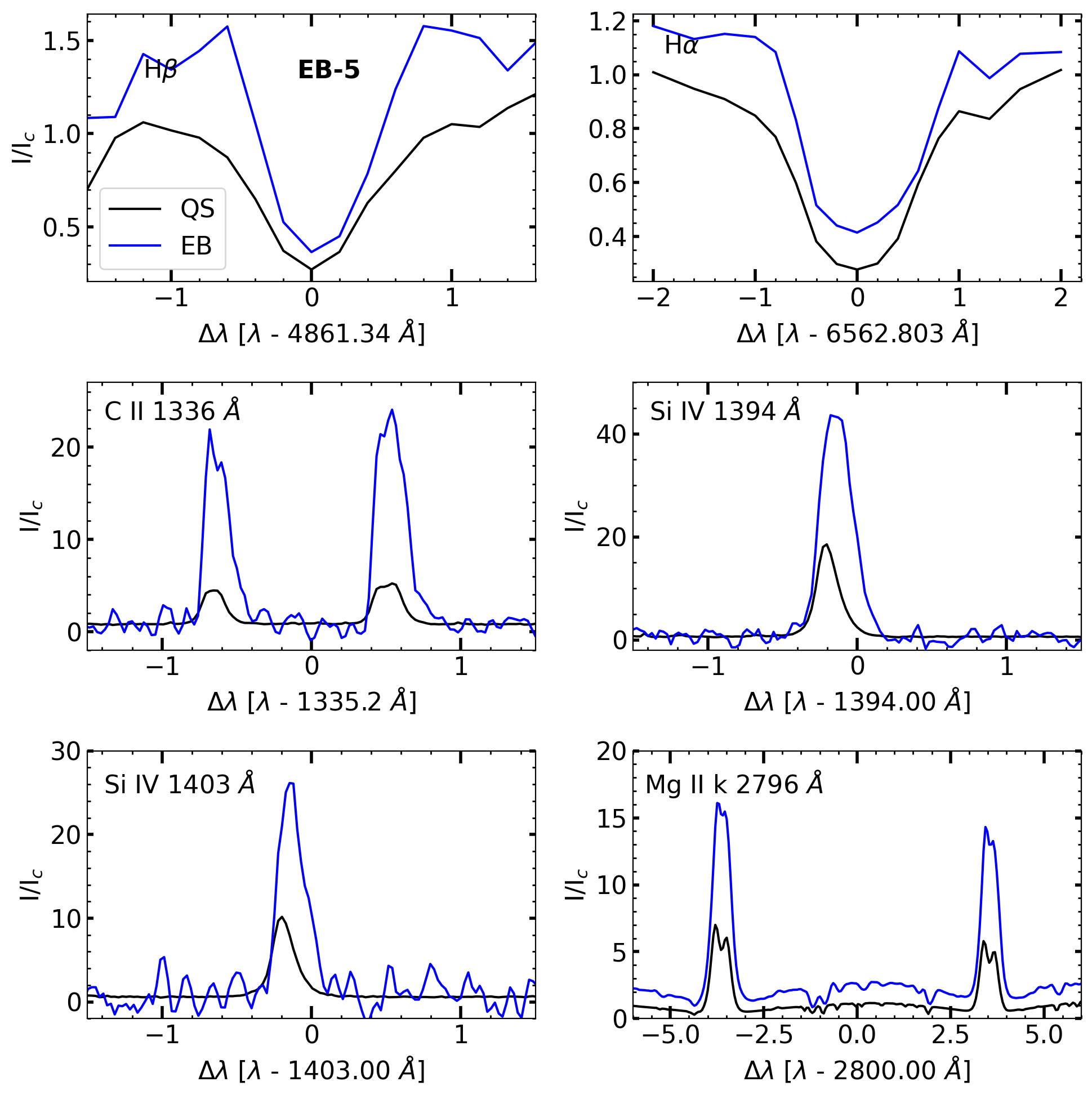}
	\includegraphics[width=85mm]{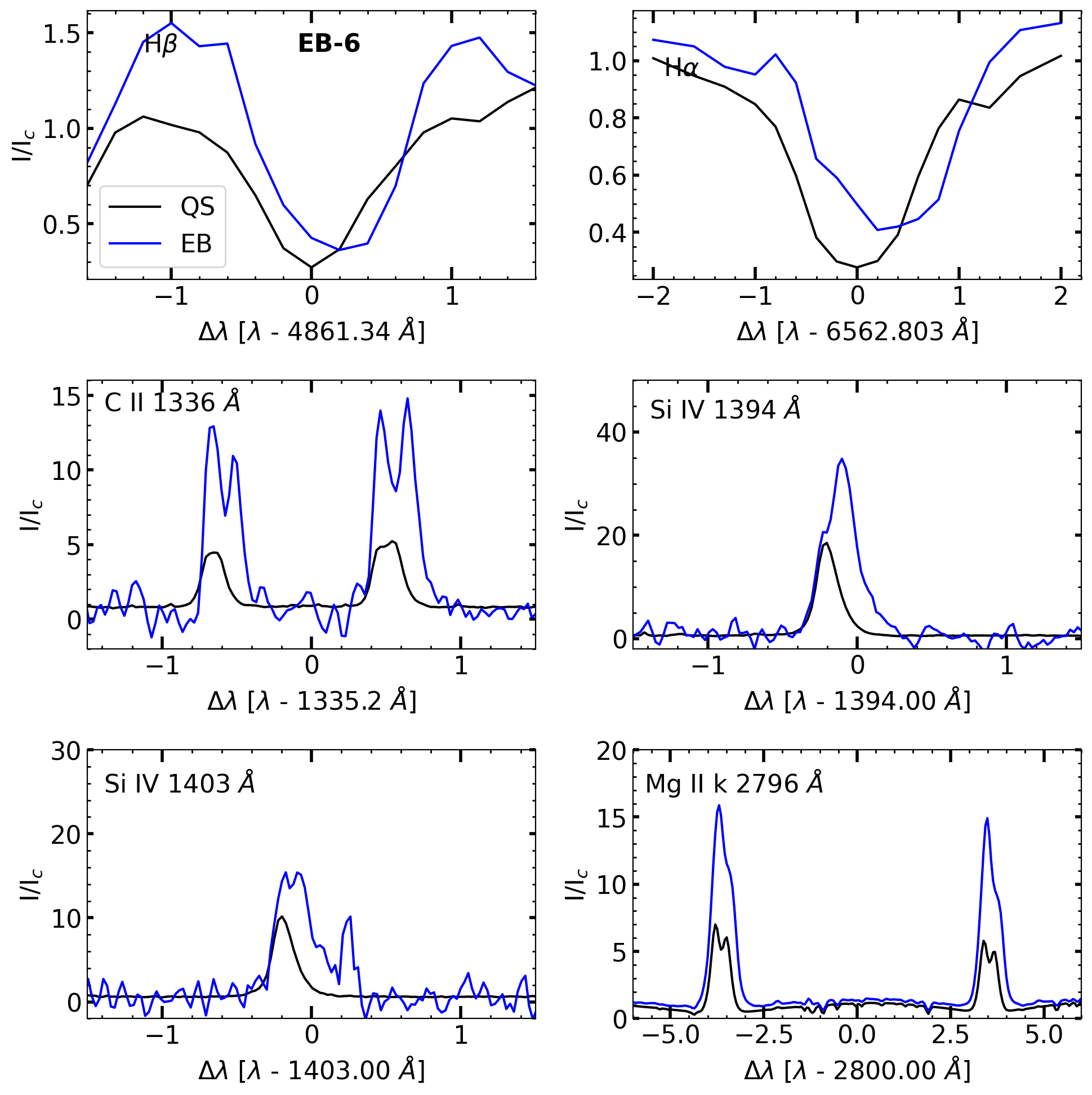}
	\caption{Spectral profiles for the two EBs events (labeled as 5 and 6 in the upper panel of Figure\ref{Fig1}). The spectral profiles include H$\alpha$, H$\beta$, C\,\textsc{ii}~1336~\AA, Si\,\textsc{iv}~1394~\AA\ and 1403~\AA, and Mg\,\textsc{ii} h\&k lines.}
	\label{Fig7}
\end{figure*}

Simultaneous enhanced emission in transition-region spectral lines, together with increased wing emission in chromospheric lines associated with EBs (Figure~\ref{Fig7}) and the presence of opposite magnetic polarities, supports the interpretation that EBs can extend to transition-region temperatures. This is consistent with earlier coordinated chromospheric and transition-region studies that reported EB signatures in transition-region diagnostics (e.g., \citealt{2011ApJ...736...71W,2013ApJ...779..125N,2015ApJ...812...11V,2020A&A...633A..58O}), suggesting that at least a subset of EBs can heat plasma to higher temperatures than previously thought. In contrast, the QSEBs analyzed in this study exhibit different behavior. Although some QSEBs show clear wing enhancement in chromospheric lines, only a subset (e.g., QSEB-3 and QSEB-4; Figure~\ref{FigA3}) display co-spatial brightenings in the IRIS SJI 1400~\AA\ channel. This is consistent with the findings of \cite{2024A&A...689A.156B}, who, using coordinated SST H$\beta$ and IRIS Si,\textsc{iv} observations, reported that some QSEBs can also be heated to transition-region temperatures. They further identified cases where QSEBs were simultaneously observed by the IRIS slit and SST, providing stronger evidence that a fraction of QSEBs may indeed reach transition-region conditions as the SJI 1400~\AA\ emission is not exclusively due to Si,\textsc{iv}, but can also include contributions from UV continuum and lower-temperature chromospheric emission \citep{2014SoPh..289.2733D}.

Furthermore, coordinated observations in coronal passbands show that both EBs and QSEBs are not detected in hotter coronal channels such as AIA 131~\AA, 171~\AA, 193~\AA, 211~\AA, and 335~\AA. This absence suggests that these events do not heat plasma to coronal temperatures, which is consistent with several earlier observational studies reporting a lack of coronal signatures associated with EBs (e.g., \citealt{2013ApJ...774...32V,2020A&A...633A..58O}). These studies collectively indicate that, although EBs are signatures of magnetic reconnection, the associated heating is generally confined to the lower chromosphere or, in some cases, extends only up to the transition region. Alternatively, their emission may be obscured by overlying magnetic canopy structures. This interpretation has been discussed in earlier works, where it was suggested that canopy fields can limit or mask the upward propagation of heating signatures from lower atmospheric layers (e.g., \citealt{2011ApJ...736...71W,2013JPhCS.440a2007R,2015ApJ...812...11V}). The latter scenario is more likely in active regions, i.e., for EBs where stronger and more complex magnetic field configurations are present, leading to enhanced canopy coverage. This interpretation is further supported by our observations: although EB signatures are clearly seen in the wings of H$\alpha$, H$\beta$, Ca,\textsc{ii}~8542~\AA, and Ca,\textsc{ii} H \& K, as well as in the AIA 1600~\AA\ and 1700~\AA\ channels, they are absent in IRIS SJI Mg,\textsc{ii} k and Si,\textsc{iv} 1400~\AA\ images. These diagnostics sample the upper chromosphere and transition region and are strongly influenced by overlying magnetic canopy structures (see the right panel of Figure~\ref{Fig4} and Figure~\ref{FigA1}). 

\subsection{Vertical Stratification of the physical parameters above the EBs/QSEBs obtained utlising STiC inversion code} \label{ssec3.3}

We employed the STiC inversion code to derive the thermodynamic structure of the solar chromosphere at the locations of EBs/QSEBs. STiC is a parallelized inversion code (MPI-based), developed to recover the vertical stratification of atmospheric parameters, making it well-suited for tracking the evolution of small-scale chromospheric phenomena. The code builds upon a modified version of the RH radiative transfer code \citep{2001ApJ...557..389U}, and applies cubic Bézier solvers to integrate the polarized radiative transfer equation \citep{2013ApJ...764...33D}. It operates under non-LTE conditions with statistical equilibrium and is capable of fitting several spectral lines simultaneously. In addition, STiC incorporates an efficient approximation for partial redistribution effects \citep{2012A&A...543A.109L}. Each pixel is treated under a plane-parallel assumption (1.5D approach), and the LTE equation of state is adopted from the Spectroscopy Made Easy (SME) package \citep{2017A&A...597A..16P}.

The inversion is carried out by iteratively modifying physical parameters, including temperature ($T$), line-of-sight velocity ($V_{\rm LOS}$), and microturbulent velocity ($V_{\rm turb}$), until the difference between the synthetic and observed spectra, measured by a $\chi^2$ merit function, is minimized. The resulting stratifications are referenced to the logarithmic optical depth at 500 nm ($\log \tau_{500}$). Gas pressure and density are subsequently derived assuming hydrostatic equilibrium.

Perturbations to the initial model are introduced through nodes defined along the $\log \tau_{500}$ axis, which are interpolated across the full depth grid. For this study, we inverted the complete Stokes profiles of the Ca~\textsc{ii} 8542~\AA\ line, adopting the Fontenla–Avrett–Loeser quiet-Sun model (FAL-C; \cite{1993ApJ...406..319F}) as the starting atmosphere. A six-level Ca~\textsc{ii} atomic model was used, with five nodes assigned to temperature ($T$), three nodes to line-of-sight velocity ($V_{\rm LOS}$) and microturbulent velocity ($V_{\rm turb}$), and two nodes to each magnetic field component. The Ca~\textsc{ii} 8542~\AA\ line core was found to be most sensitive to conditions near $\log \tau_{500} = -4.8$.

The inferred physical parameters, including temperature ($T$), line-of-sight velocity ($V_{\rm LOS}$), line-of-sight magnetic field ($B_{\rm LOS}$), and microturbulent velocity ($V_{\rm turb}$), obtained from STiC inversions of four EB events are presented in Figure~\ref{Fig_STiC_EBs}. A similar analysis was carried out for QSEBs; however, only $T$, $V_{\rm LOS}$, and $V_{\rm turb}$ are shown in Figure~\ref{Fig_STiC_QSEBs}, as the Ca \textsc{ii} 8542~\AA\ observations for Dataset~2 were acquired in spectroscopic mode.

For EBs, we find an average temperature enhancement of $\sim$ 1700 K in the chromospheric layers spanning $\log_{5000}\tau \approx -3.0$ to $-5.0$. In addition to this temperature increase, significant variations in $v_{\text{los}}$, $B_{\text{los}}$, and $v_{\text{turb}}$ are observed across these heights, consistent with localized energy release and dynamic plasma motions associated with magnetic reconnection in the lower chromosphere. The vertical lines shown in Figure \ref{Fig_STiC_EBs} represent the 1$\sigma$ uncertainties of the derived thermodynamic parameters. These uncertainties were estimated using a Monte Carlo approach by perturbing the model atmosphere and repeating the inversions 500 times. The standard deviation of the resulting distributions was adopted as the 1$\sigma$ uncertainty.

Similar temperature enhancements have been reported in several previous studies of EBs. For example, \cite{2002ApJ...575..506G} estimated temperature enhancements of $\sim$ 2000 K within the radiating volume, while \cite{Fang.et.al.06}, using semi-empirical EB models, inferred additional heating of $\sim$ 600 to 1300 K near the temperature minimum region. \cite{Berlicki&Heinzel.14} reported maximum temperature increases ranging from $\sim$ 1100 to $\sim$ 5500 K, whereas \cite{Hong.et.al.17}) inferred temperature enhancements up to $\sim$ 2300 K. More recently, \cite{da.Silva.Santos.et.al.25} reported temperature increases spanning $\sim$ 100 to 2000 K. These values indicate that moderate heating of the lower atmosphere is a common characteristic of EBs. Our results are also broadly consistent with inversion-based studies by \cite{2019A&A...627A.101V}, who, using Mg \textsc{ii} H \& K, Ca \textsc{ii}~8542~\AA\, and Ca \textsc{ii} H \& K diagnostics, reported average temperature enhancements of a few thousand kelvin near the classical temperature minimum, together with localized peak temperatures reaching $\sim$ 10,000 to 15,000 K. They further noted that including Mg II diagnostics generally reduces inferred temperature enhancements to below $\sim$ 8000 K, highlighting the sensitivity of inversion results to the adopted spectral diagnostics. Other studies suggest that substantially higher temperatures may occasionally be achieved at deeper atmospheric layers (e.g., \cite{Leenaarts.et.al.25}), emphasizing the dependence of inferred temperatures on the selected diagnostics, inversion approach, and atmospheric assumptions. Compared to these studies, the lower temperature enhancement inferred here may partly reflect differences in diagnostics, inversion setup, and the fact that the Ca \textsc{ii}~8542~\AA\ line primarily samples the lower chromosphere. Nevertheless, our results confirm localized heating associated with EBs and support the interpretation that these events are driven by magnetic reconnection.

In contrast, we do not find significant temperature enhancements in QSEBs. The absence of clear thermodynamic signatures may be attributed to the relatively coarse wavelength sampling and limited spatial resolution of the Ca \textsc{ii}~8542~\AA\ observations. This interpretation is supported by the coordinated observations (see Figure~\ref{Fig6}), where QSEBs are clearly identifiable in H$\alpha$ and H$\beta$ but remain ambiguous in the corresponding Ca \textsc{ii}~8542~\AA\ images.

\begin{figure}
	\centering
	\includegraphics[width=83.7mm]{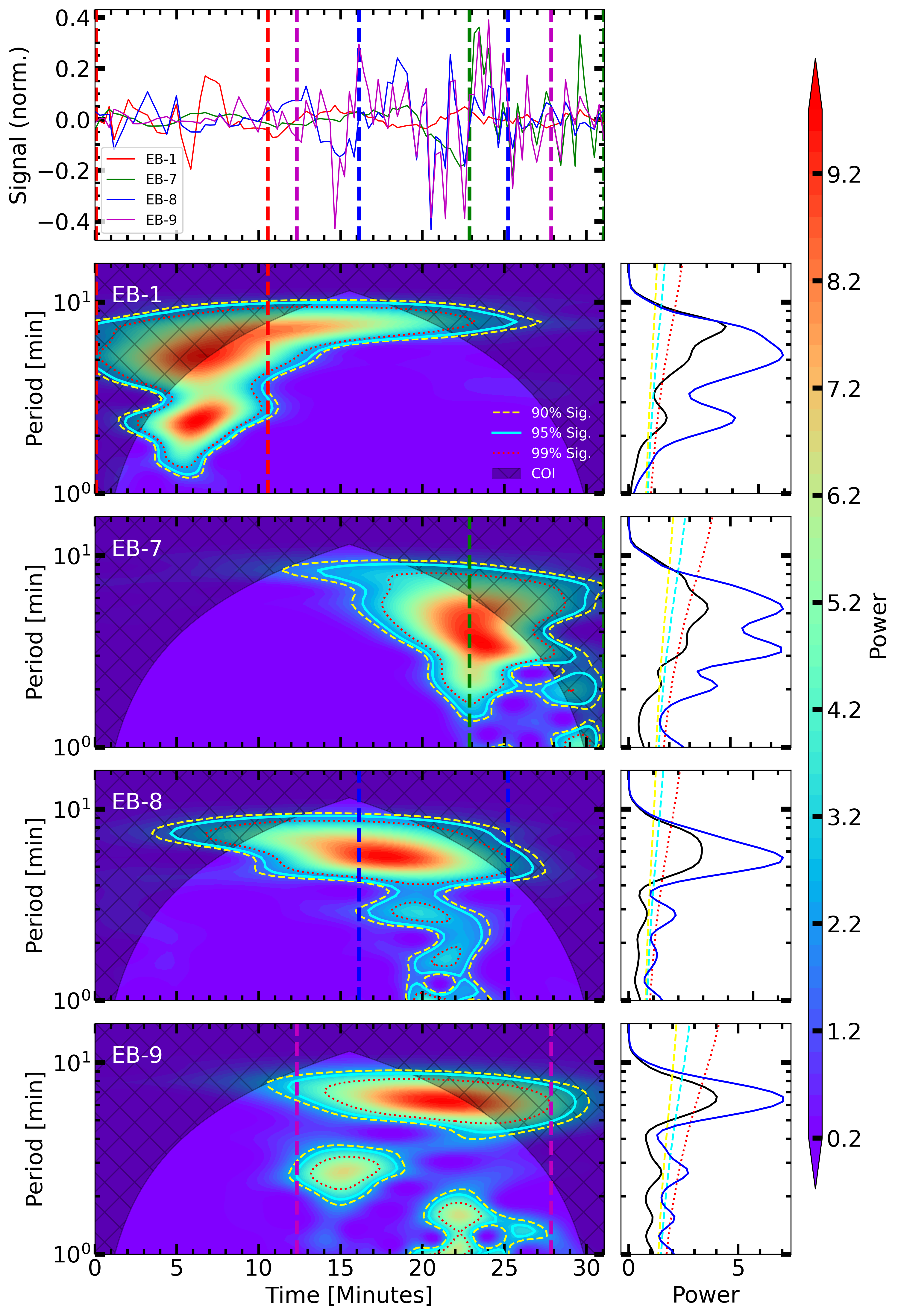}
	\caption{Wavelet analysis of the average intensity variations of the H$\beta$ spectral line over the wavelength range $-0.71$ to $-0.44$~\AA\ for four EB events (labeled 1, 7, 8, and 9 in the upper panel of Figure~\ref{Fig1}). The vertical dashed magenta lines indicate the lifetimes of the EBs, while the contours denote different significance levels. In the right panels, the black curves show the global wavelet power spectra, and the blue curves represent the power spectra averaged over the EB lifetimes only.}
	\label{Fig9}
\end{figure}

\subsection{Rate of Episodic Heating/Reconnection in the Lower Chromosphere}\label{ssec3.4}

To investigate the rate of episodic heating/reconnection in the lower chromosphere, we selected a subset of isolated EBs (labeled 1, 7, 8, and 9) that persisted for a well-defined duration at fixed locations and did not reoccur at the same positions during the observation period. Such isolated events provide ideal conditions for examining their localized temporal evolution without interference from recurrent activity. We performed wavelet analysis \citep{1998BAMS...79...61T} on the average intensity variations of the linearly interpolated H$\beta$ spectral line over the wavelength range $-0.71$ to $-0.44$~\AA, where EBs exhibit maximum sensitivity. Prior to the wavelet analysis, spline interpolation was applied to the spectral profiles to improve the wavelength sampling. Additionally, the slow background variations were removed by subtracting a 7th-order polynomial fit from the time series. This detrending was applied to suppress long-term trends and enhance sensitivity to transient periodicities associated with EB activity. Representative examples of this analysis are presented in Figure~\ref{Fig9}. The significance levels in the wavelet analysis were initially estimated assuming a white-noise background spectrum. We additionally repeated the analysis using a red-noise background spectrum to assess the robustness of the inferred periodicities. The dominant periods and associated power enhancements remained broadly consistent under both assumptions, indicating that the principal periodicities reported here are robust regardless of the choice of background noise model.

It should be noted that wavelet analysis was not applied to Doppler velocity fluctuations, as the line profiles are significantly distorted during these events, preventing reliable velocity estimation. A similar analysis was not performed for QSEBs because their comparatively short lifetimes provide insufficient temporal coverage to robustly characterize periodic behavior or recurrent heating signatures on timescales comparable to those inferred for EBs. Moreover, the much shorter duration of QSEBs may also suggest that these events do not exhibit periodic behavior at all, or at least not on the same timescales as EBs.

To assess the impact of linear interpolation used to account for missing frames on periodicity estimation, we additionally applied the Lomb–Scargle periodogram to the original irregularly sampled time series. The dominant periods obtained were broadly consistent with those inferred from the global wavelet spectra, indicating that interpolation does not significantly affect the principal periodicities discussed here.

The wavelet analysis reveal that although multiple dominant periodicities are present, a period of approximately 6--7 minutes appears consistently in a subset of the analyzed EBs. This periodic behavior is most prominent during the EB lifetimes, as indicated by the enhanced power in the time-averaged spectra (see Figure~\ref{Fig9}). The presence of this preferred timescale suggests that the reconnection process associated with some EBs occurs in a quasi-periodic manner, possibly driven by repetitive magnetic flux interactions or modulation by underlying chromospheric oscillations.

\subsection{Possible association of some of the spicule origination with the EBs/QSEBs}\label{ssec3.5}

\begin{figure*}
	\centering
	\includegraphics[width=150mm]{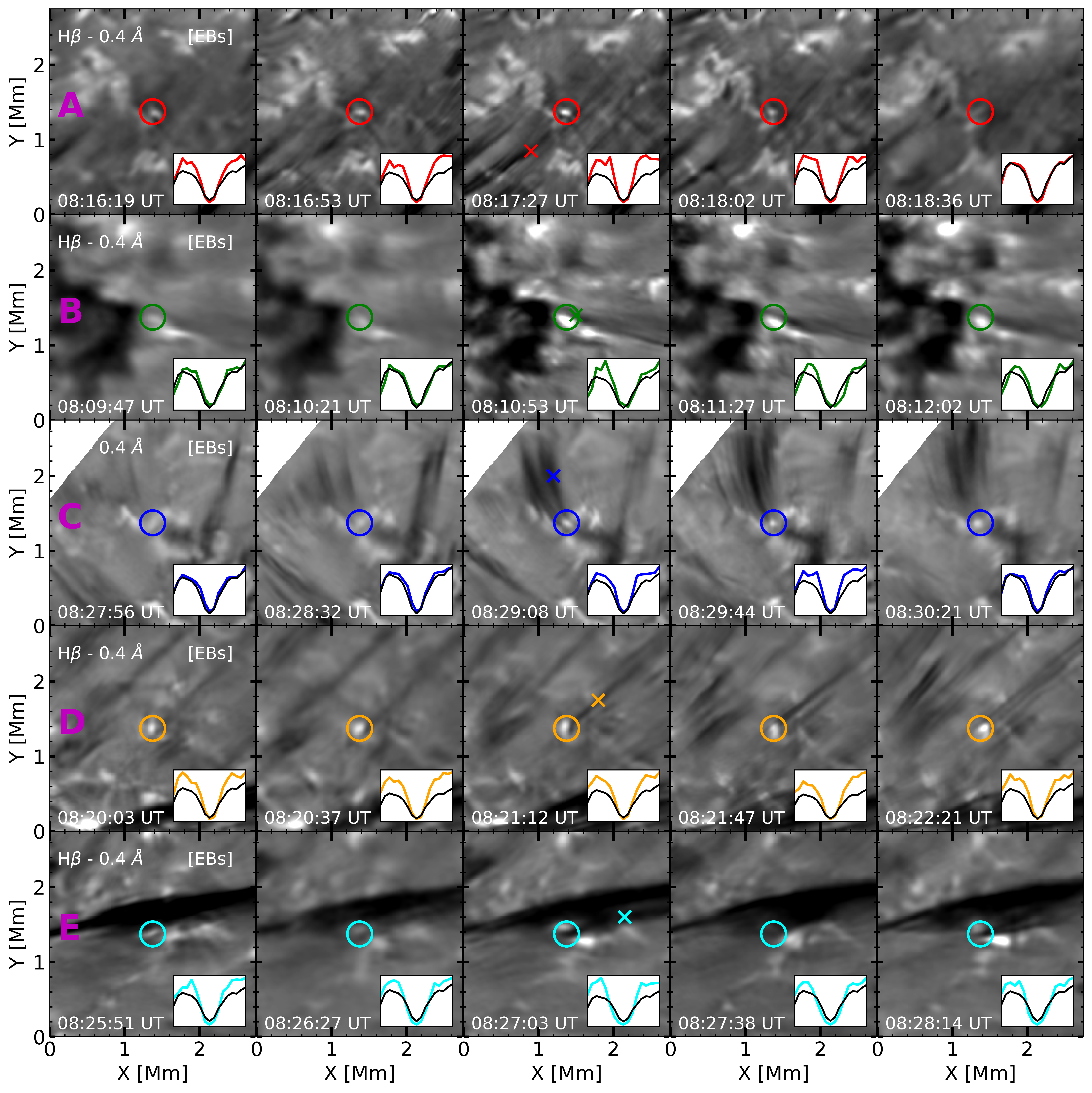}
    \caption{Representative examples showing image sequences that illustrate the temporal evolution of five EB-associated spicule events (rows A--E) observed in the H$\beta$ $-0.4$ \AA\ wing. Colored circles mark the locations of the EBs, while the corresponding crosses indicate the positions of the associated spicules, whose H$\beta$ spectral profiles are shown in the left panel of Figure~\ref{Fig13}. The displayed images are separated by two intermediate frames. The inset profiles display the normalized H$\beta$ spectra extracted at the EB locations, together with the average quiet-Sun profile (black curve). The EB spectra exhibit the characteristic wing-enhanced emission signatures. The image sequences suggest a temporal and spatial association between some EBs and appearance of spicules whose apparent footpoints coincide with the EB locations.}
\label{Fig11}	
\end{figure*}

\begin{figure*}
	\centering
	\includegraphics[width=150mm]{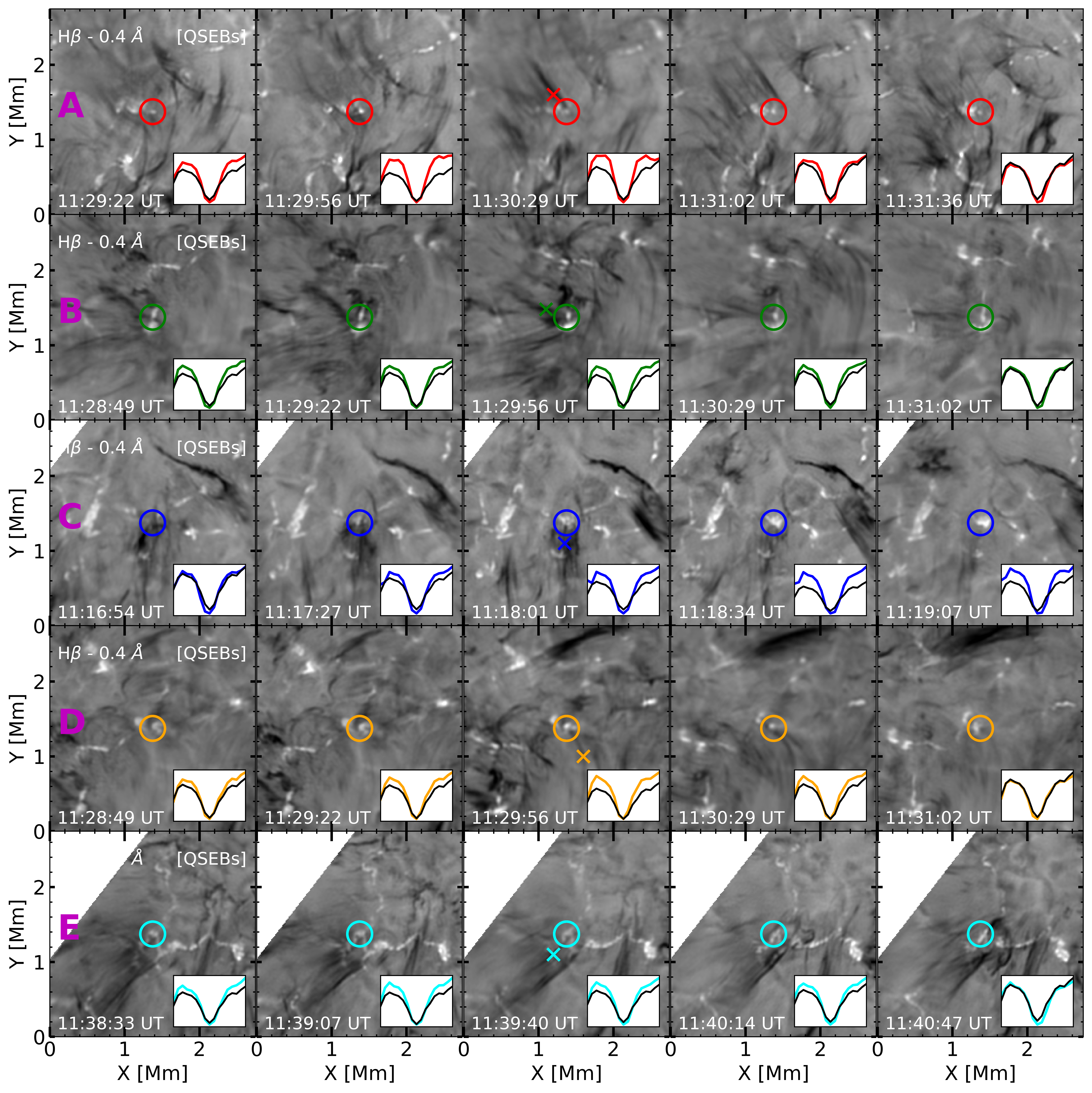}
    \caption{Same as Figure~\ref{Fig11}, but for five QSEB-associated spicule events observed in the H$\beta$ $-0.4$ \AA\ wing. The H$\beta$ spectral profiles of the spicule locations marked by crosses are shown in the right panel of Figure~\ref{Fig13}. The inset profiles correspond to the QSEB locations and display the characteristic, although generally weaker, wing-enhanced emission signatures compared to classical EBs.}
	\label{Fig12}	
\end{figure*}

Finally, we investigated the possible connection between the origin of spicules and the occurrence of EBs/QSEBs in the solar atmosphere. Our analysis focused on identifying cases in which the apparent footpoints of jet-like chromospheric structures were spatially associated with EBs/QSEBs observed in the H$\beta$ scans from Dataset~1 and Dataset~2. Figures~\ref{Fig11} and \ref{Fig12} present five representative examples of EB-associated and QSEB-associated spicule events, respectively.

The image sequences reveal a close spatial and temporal association between some EBs and spicule-like events. In some cases, such as events A and E in Figure~\ref{Fig11}, the EB first appears as a localized wing brightening, followed a few frames later by the origin of an elongated absorption feature whose apparent footpoint coincides with the EB location. Events B and C show a nearly simultaneous appearance of the EB and the spicule-like feature. In event D, both the EB and the spicule-like structure are already present throughout the displayed image sequence, making it difficult to determine the temporal ordering of the two phenomena. The observations therefore suggest a mixed behavior, with some events showing a clear sequence in which the EB precedes the jet-like feature, while others exhibit a simultaneous presence of both signatures.

A similar behavior is found for the QSEB-associated events shown in Figure~\ref{Fig12}. In particular, events A, B, and D suggest that the spicule-like feature appears a few frames after the QSEB brightening becomes visible. In contrast, event C shows a nearly simultaneous appearance of the QSEB and the spicule-like feature, while in event E both the QSEB and the associated jet-like structure are already present in the displayed image sequence.

\begin{figure*}
	\centering
	\includegraphics[width=80mm]{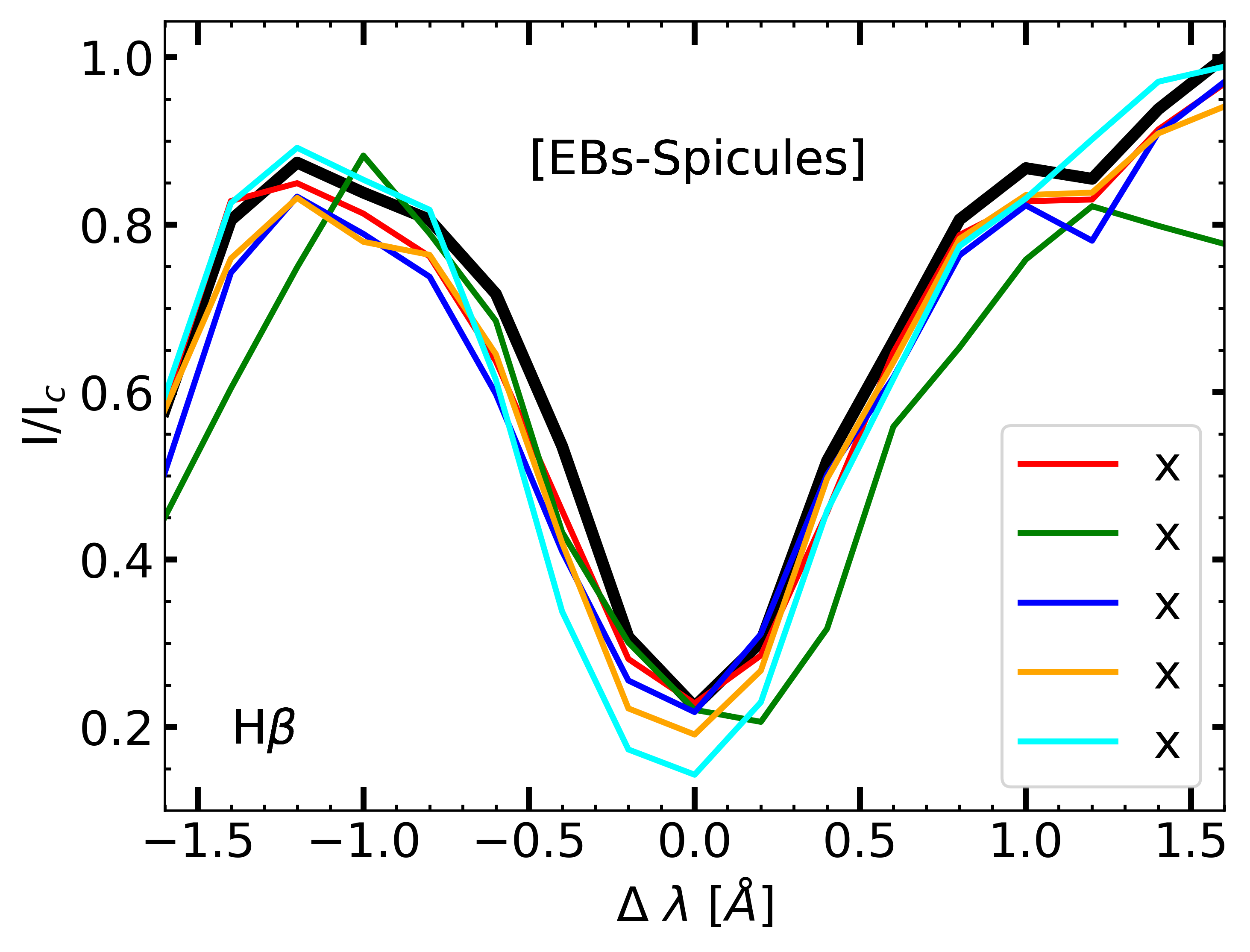}
	\includegraphics[width=80mm]{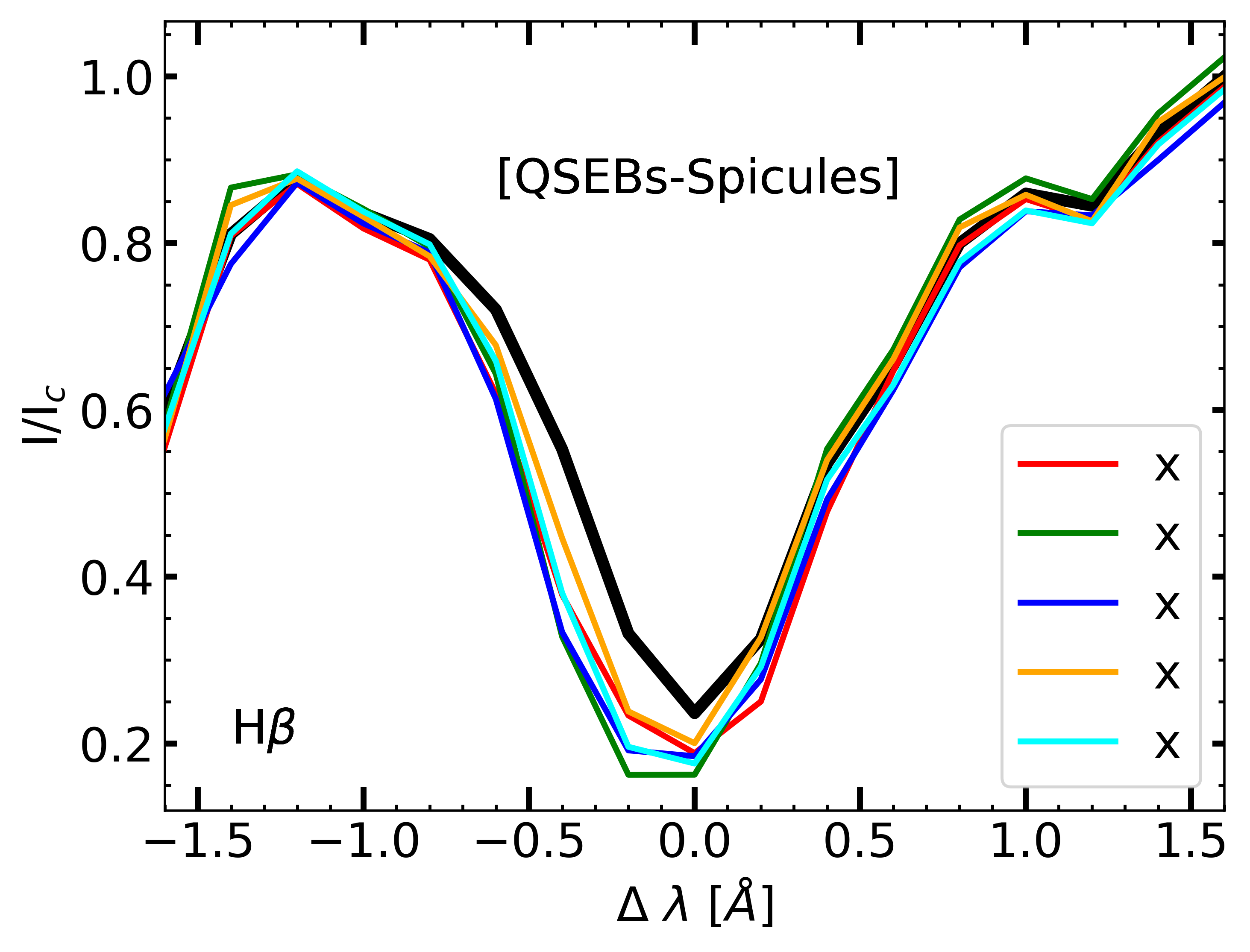}
    \caption{Normalized H$\beta$ spectral profiles of the spicule associated with the EB and QSEB events shown in Figures~\ref{Fig11} and \ref{Fig12}. The left panel shows the spectra extracted at the locations marked by crosses in the EB-associated events, while the right panel shows the corresponding spectra for the QSEB-associated events. The black curves represent the average quiet-Sun profile. The red and blue-shifted absorption signatures are characteristic of RREs and RBEs, which are commonly interpreted as on-disk counterparts of spicules.}
	\label{Fig13}	
\end{figure*}

To verify that these jet-like structures are consistent with spicules, we examined the H$\beta$ spectra at the locations marked by the crosses in Figures~\ref{Fig11} and \ref{Fig12}. The corresponding spectral profiles, shown in Figure~\ref{Fig13}, exhibit pronounced red or blue-shifted absorption signatures relative to the average profile. Such signatures are characteristic of Rapid Red-shifted Excursions (RREs) and Rapid Blue-shifted Excursions (RBEs), which are widely regarded as the on-disk counterparts of spicules \citep{2009ApJ...705..272R,Bose.et.al.21,2024ApJ...970..179C}. Similar spectral behavior is observed for both the EB-associated and QSEB-associated events presented here.

These observations are qualitatively consistent with the recent results of \citet{2025A&A...697A.180S}, who reported both cases in which QSEBs precede the appearance of spicule-like structures and cases where QSEBs and spicule-like features are observed simultaneously. They suggested that magnetic reconnection occurring in the lower atmosphere may contribute to the generation of at least a subset of spicules. While our observations do not establish a causal relationship, they provide additional examples supporting the possibility that some spicules may originate from magnetic reconnection events that manifest as EBs/QSEBs in the lower solar atmosphere.

However, it is important to note that EBs/QSEBs are not always observed at spicule footpoints. Many spicules are likely generated by alternative mechanisms, such as shock-wave-driven processes, as demonstrated in earlier studies \citep{2004Natur.430..536D,2006ApJ...647L..73H,2007ApJ...660L.169R,2022NatPh..18..595D,Chaurasiya.et.al.26}. Therefore, our results suggest that multiple formation mechanisms may coexist, with magnetic reconnection associated with EBs/QSEBs representing one possible pathway among several contributing processes.

\section{Summary \& Conclusion}\label{sec5}

In this study, we analyzed and compared the properties of EBs and QSEBs using coordinated multiwavelength observations from SST, IRIS, and SDO. By combining high-resolution observations with statistical classification and non-LTE inversions, we investigated their spectral signatures, thermal structure, temporal behavior, and possible connection to spicule formation. Our results show that, although both EBs and QSEBs exhibit clear chromospheric intensity enhancements, their physical properties differ significantly. EBs are associated with stronger magnetic fields and frequently display opposite magnetic polarities, consistent with magnetic reconnection. A subset of EBs also shows signatures in transition-region diagnostics, indicating heating to higher temperatures, while neither EBs nor QSEBs are detected in coronal passbands, suggesting that their heating remains confined below coronal temperatures or is obscured by magnetic canopy structures. In contrast, QSEBs appear predominantly confined to the chromosphere, lacking clear opposite magnetic polarities and showing limited transition-region signatures. STiC inversions further support this distinction, revealing temperature enhancements of $\sim$1700~K and dynamic plasma signatures for EBs, whereas QSEBs do not show clear temperature variations. This may reflect intrinsically weaker heating or observational limitations of the Ca \textsc{ii}~8542~\AA\ line. We also find evidence for quasi-periodic behavior in a subset of EBs, with a characteristic timescale of $\sim$6--7 minutes, suggesting repetitive or oscillation-driven reconnection. Additionally, some spicule footpoints are co-spatial with EBs/QSEBs, supporting a link between small-scale reconnection and chromospheric jet formation, although multiple mechanisms, including shock-driven processes, are likely involved. 

Overall, EBs and QSEBs represent related but physically distinct manifestations of small-scale energy release in the lower solar atmosphere. Although our analysis does not provide direct evidence for substantial heating associated with QSEBs, their high occurrence rates and possible connection to small-scale magnetic reconnection as found in previous studies \citep{2016A&A...592A.100R,2020A&A...641L...5J} suggest that they remain important targets for future investigations of chromospheric energy release. In addition to their potential contribution to heating, QSEBs may also play a role in the reorganization and restructuring of the magnetic field through resistive processes. Improved spatial resolution and multi-height diagnostics may help determine whether weak heating signatures associated with QSEBs contribute cumulatively to chromospheric heating and mass transport.

\begin{acknowledgments}

We thank the referee for his/her comments and suggestions. J.J. acknowledges funding support from the SERB-CRG grant (CRG/2023/007464) provided by the Anusandhan National Research Foundation, India. The Swedish 1-m Solar Telescope is operated on the island of La Palma by the Institute for Solar Physics of Stockholm University in the Spanish Observatorio del Roque de los Muchachos of the Instituto de Astrofísica de Canarias. The Swedish 1-m Solar Telescope, SST, is co-funded by the Swedish Research Council as a national research infrastructure (registration number 4.3-2021-00169). SDO is a mission for NASA Living With a Star program. The SDO/HMI data were provided by the Joint Science Operation Centre (JSOC). IRIS is a NASA small explorer mission developed and operated by LMSAL, with mission operations executed at NASA Ames Research Center and major contributions to downlink communications funded by the ESA and the Norwegian Space Center. We have made much use of NASA’s Astrophysics Data System Bibliographic Services.

\end{acknowledgments}

\appendix

\section{Additional example of the multi-wavelength observation of EBs and QSEBs in the solar atmosphere} \label{Apd1}

This section presents two additional results from the multiwavelength observations of EBs and QSEBs, covering multiple temperature regimes in the solar atmosphere. Additionally, it includes two complementary examples of EBs as observed in the Ca\,\textsc{ii} 8542~\AA\ line from MAST.

\begin{figure*}
	\centering
	\includegraphics[width=60mm]{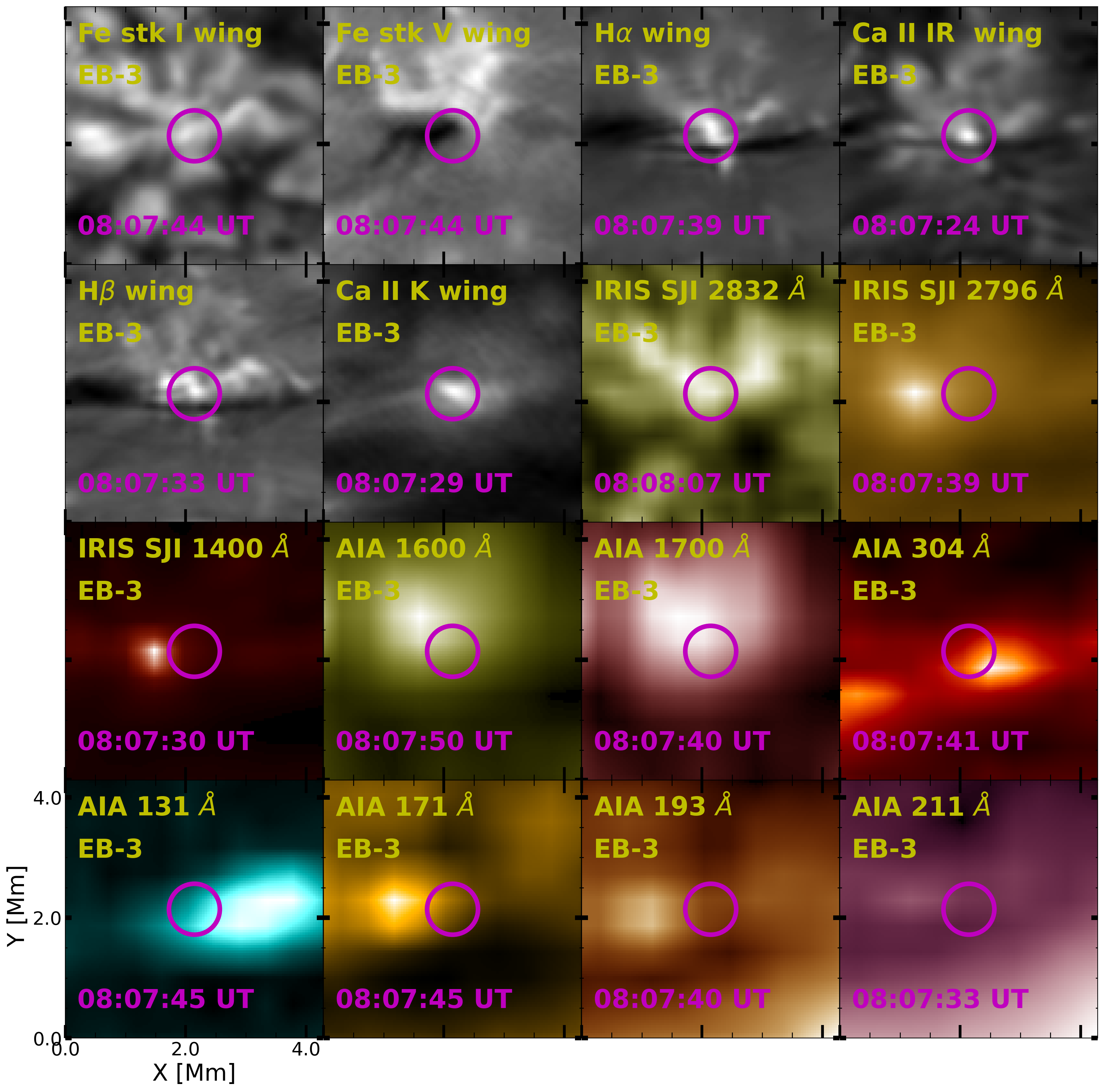}
	\includegraphics[width=60mm]{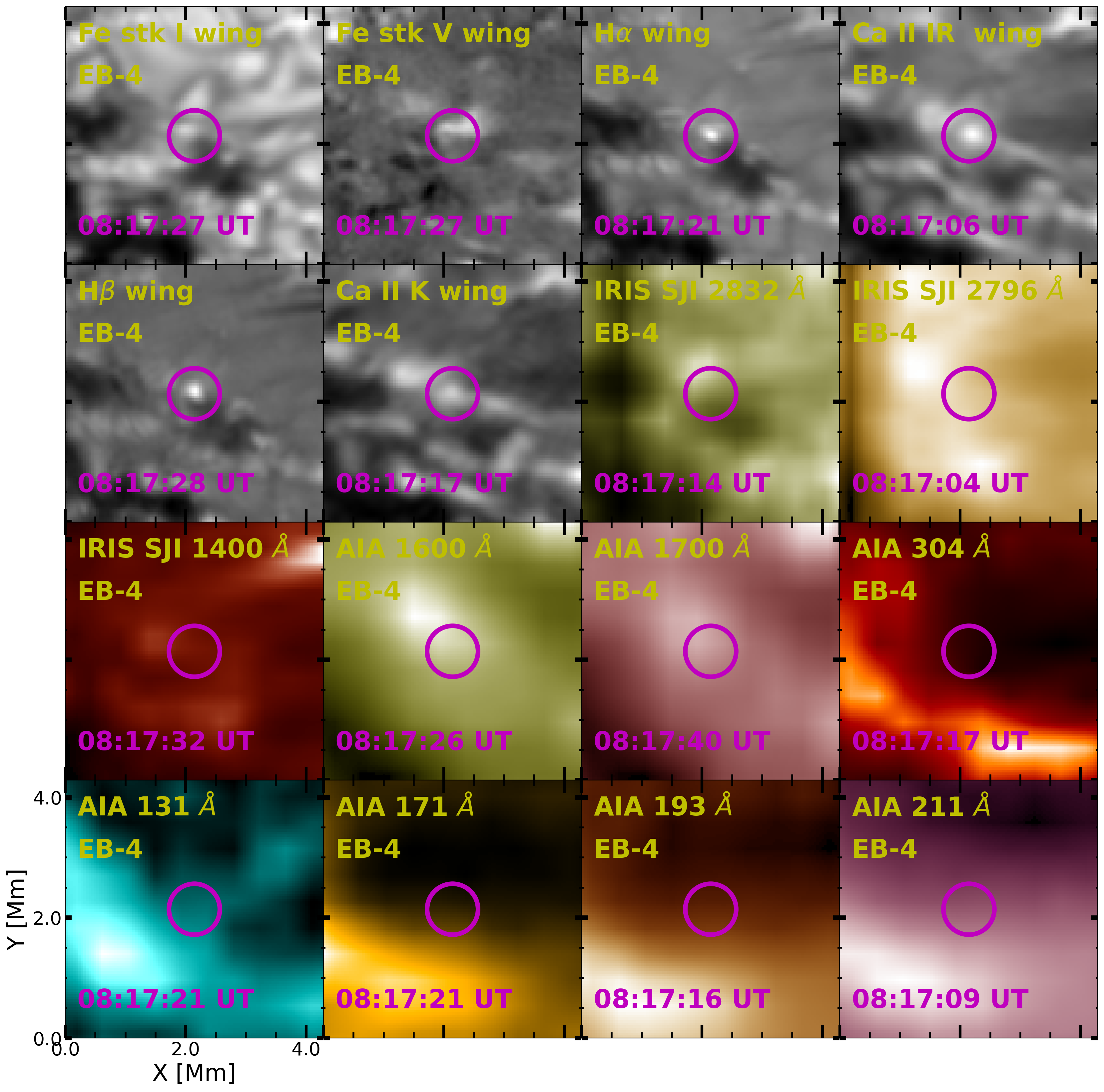}
	\includegraphics[width=60mm]{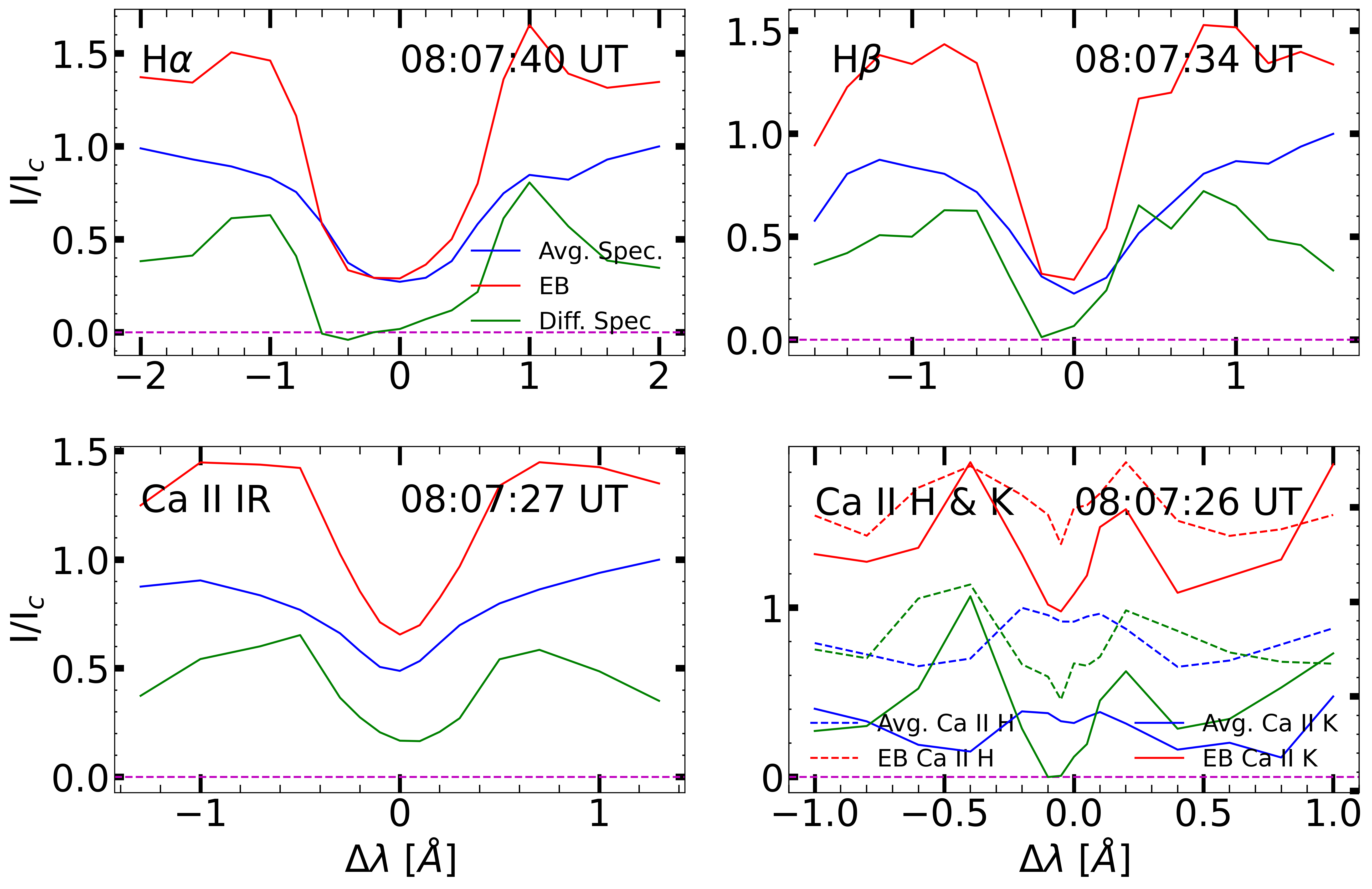}
	\includegraphics[width=60mm]{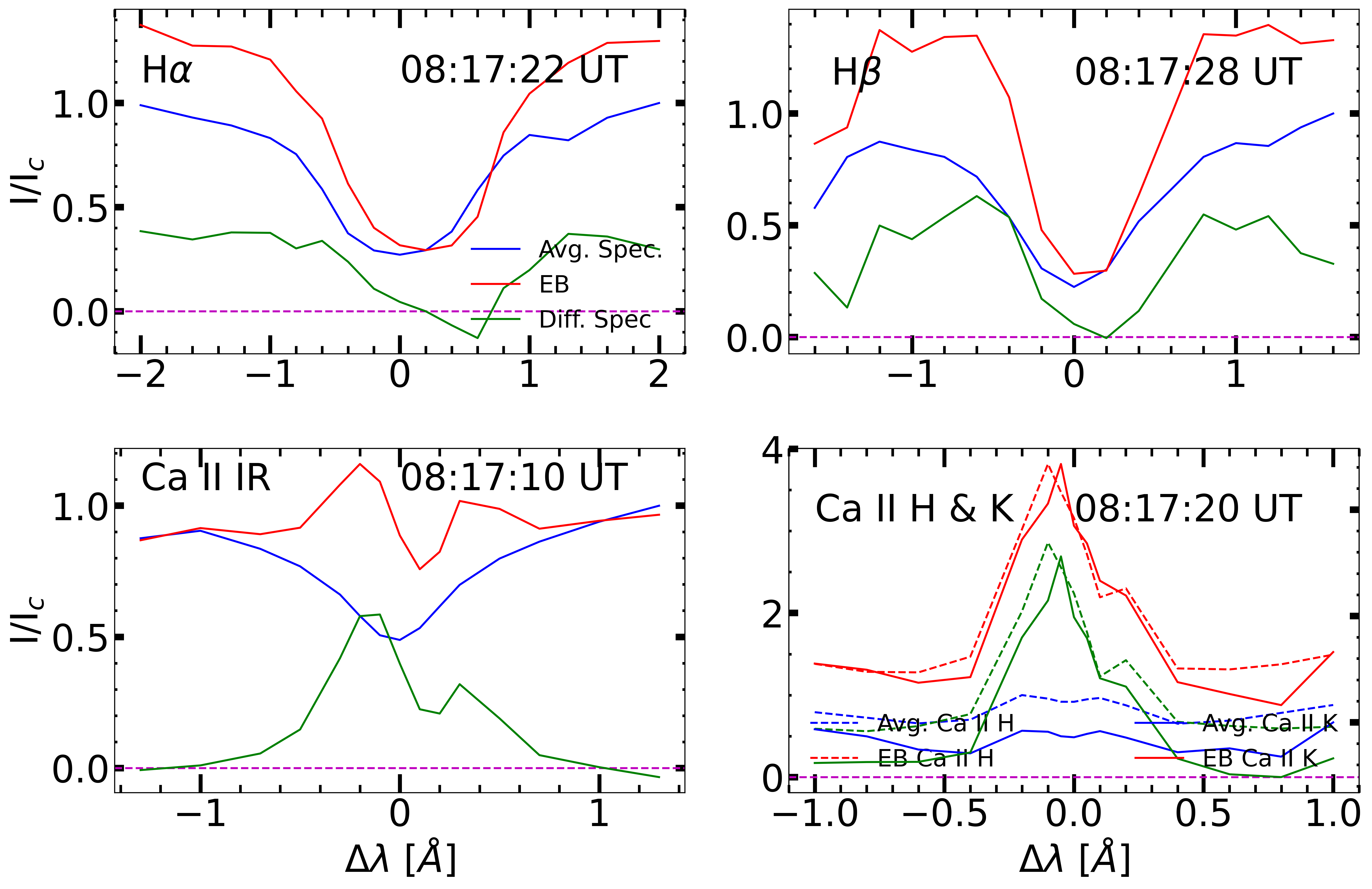}
	
	\caption{Same as Figure \ref{Fig4} but for two different EBs in the solar atmosphere
	}
	\label{FigA1}
\end{figure*}

\begin{figure*}
	\centering
	\includegraphics[width=60mm]{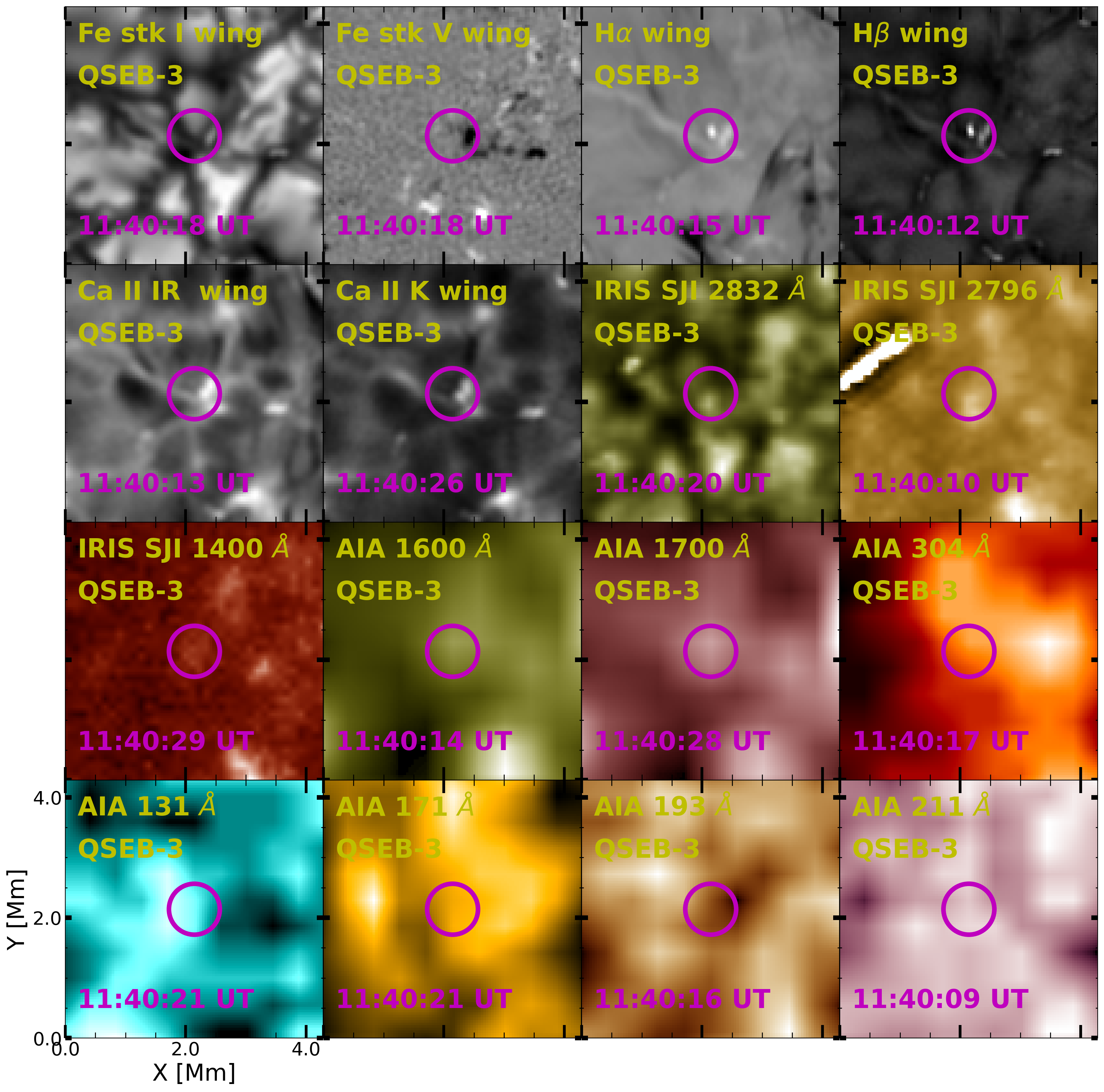}
	\includegraphics[width=60mm]{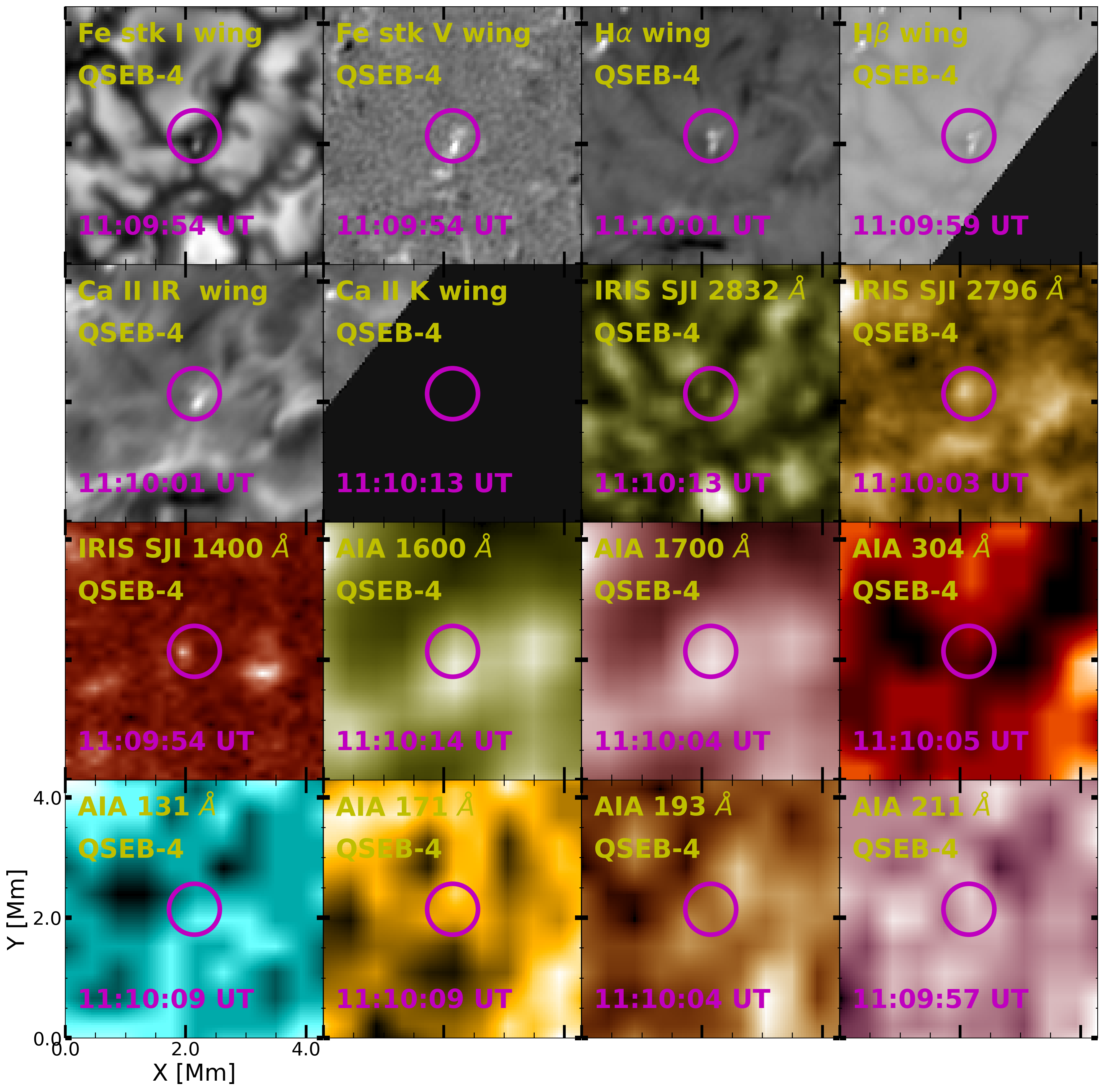}
	\includegraphics[width=60mm]{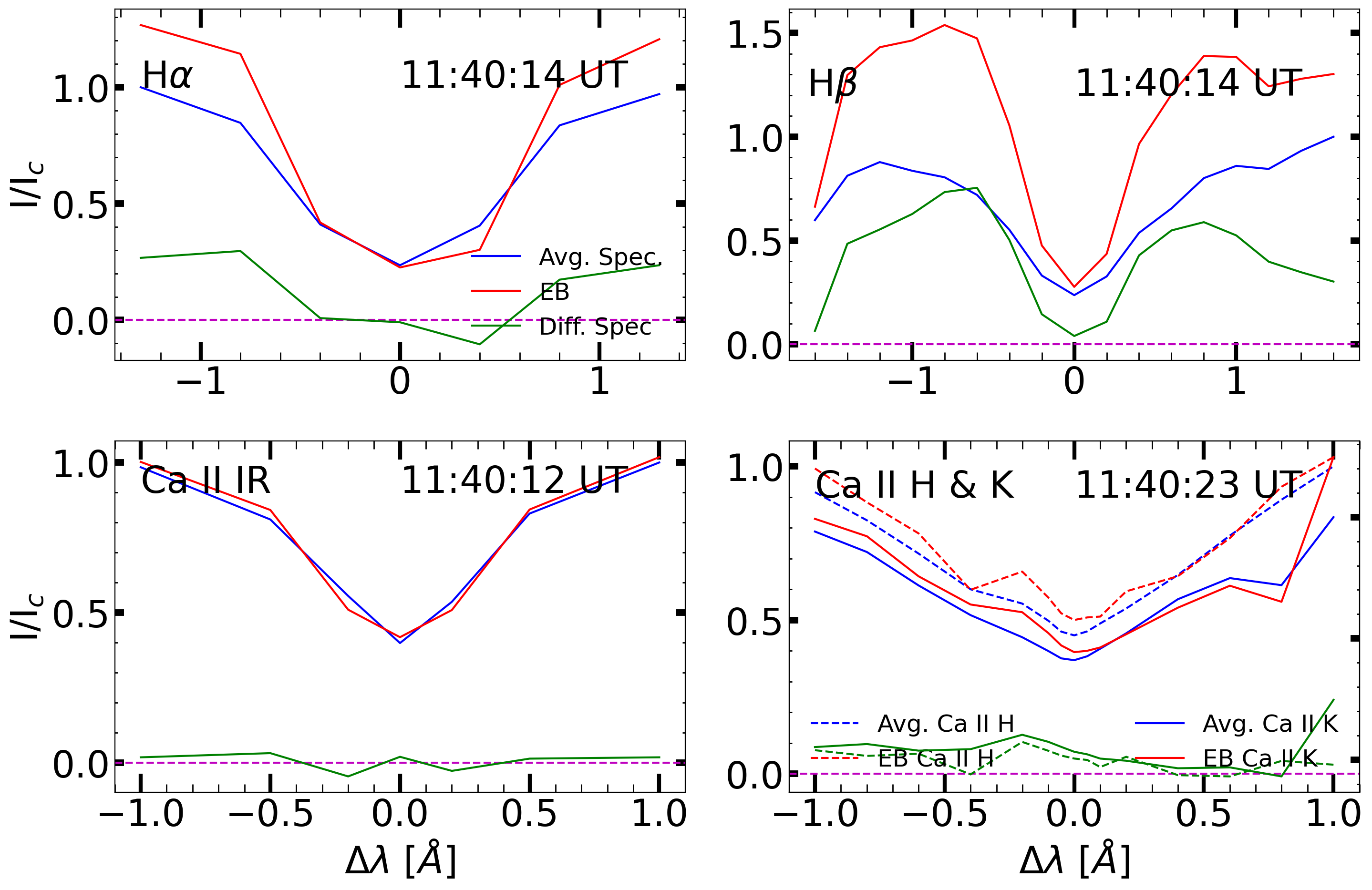}
	\includegraphics[width=60mm]{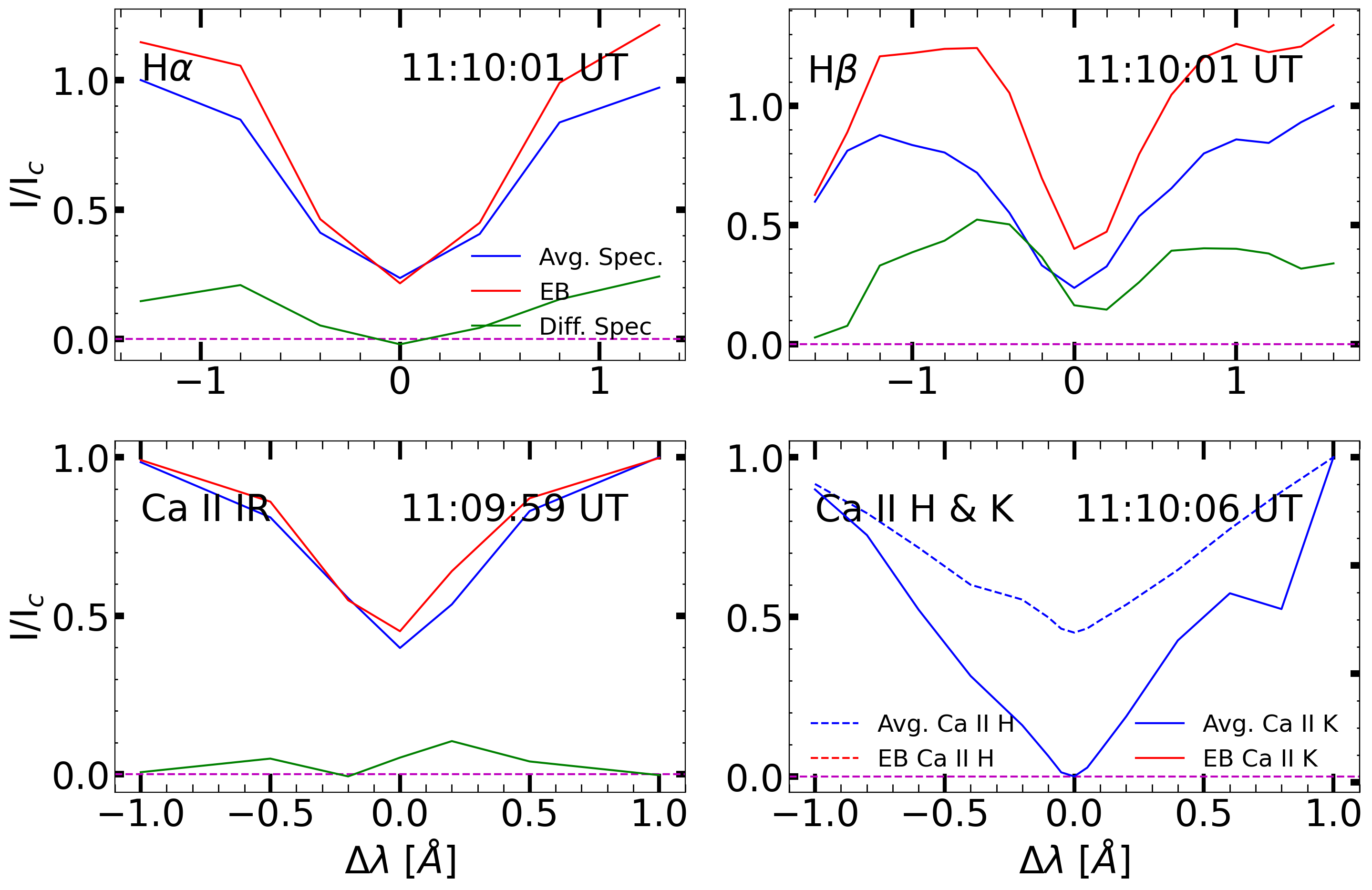}
	
	\caption{Same as Figure \ref{Fig6} but for two different QSEBs in the solar atmosphere.}
	\label{FigA3}
\end{figure*}

\begin{figure}
	\centering
	\includegraphics[width=85mm]{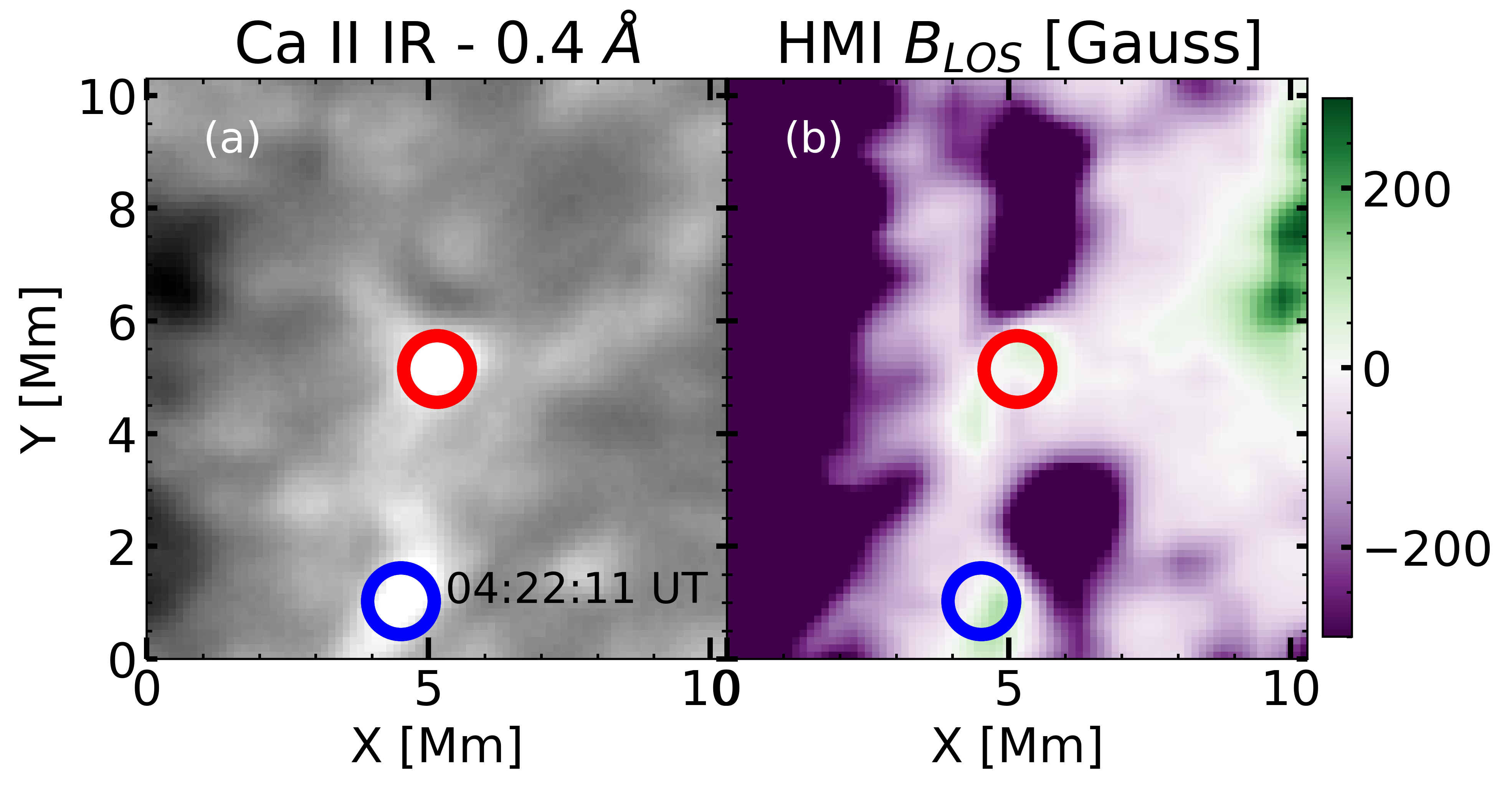}
	\includegraphics[width=85mm]{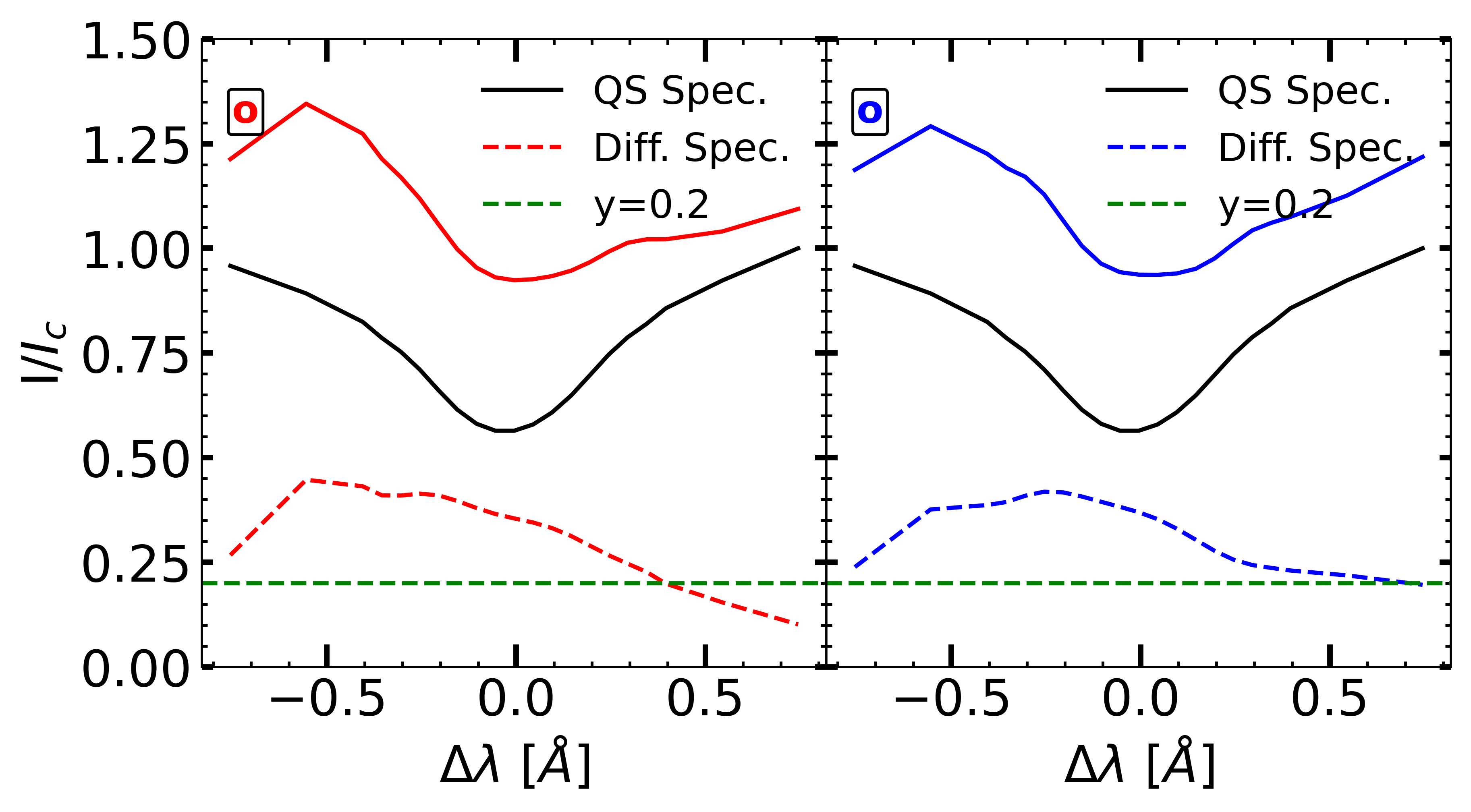}
	
	\caption{
		Same as Figure \ref{Fig5} but for two different EBs at two different locations.}
	
	\label{FigA2}
\end{figure}

\bibliography{EBs_study}{}
\bibliographystyle{aasjournalv7}



\end{document}